\documentclass[vecphys]{svmult}     % vectors in bold face
\usepackage{makeidx}         % allows index generation
\usepackage{graphicx}        % standard LaTeX graphics tool
\usepackage{epsfig}          % another way to include figures
\usepackage{multicol}        % used for the two-column index
\usepackage[bottom]{footmisc}% places footnotes at page bottom
\usepackage{amsmath,amssymb,mathrsfs}
\usepackage{ifthen}

\makeindex             % used for the subject index
\newcommand{\re}[1]{(\ref{#1})}
\newcommand{\sech}{\mathop{\mathrm{sech}}}

\newcommand{\opL}{\mathop{\mathbf{L}}}

\begin{document}
%%%%%%%%%%%%%%%%%%%%%%%%%%%%%%%%%%%%%%%%%%%%%%%%%%%%%%%%%%%%%%%%%%%%%

\def\jcmindex#1{\index{#1}}
\def\myidxeffect#1{{\bf\large #1}}

% Title
\title*{Some Recent Developments on Kink Collisions and Related Topics}
\author{
Tomasz Roma\'nczukiewicz\inst{1}
\and
Yakov Shnir\inst{2}
}
\institute{
Institute of Physics, Jagiellonian University, Krak\'ow, Poland, \texttt{trom@th.if.uj.edu.pl}
\and
BLTP, JINR, Dubna 141980, Moscow Region, Russia,~ \texttt{shnir@theor.jinr.ru}
}
\maketitle
\abstract
{
  We review recent works on modeling of dynamics of kinks in 1+1 dimensional $\phi^4$ theory and other
  related models, like sine-Gordon model or $\phi^6$ theory.
  We discuss how the spectral structure of small perturbations can affect the dynamics of non-perturbative
  states, such as kinks or oscillons.
  We describe different mechanisms, which may lead to the occurrence of the resonant structure in the kink-antikink
  collisions.
  We explain the origin of the radiation pressure mechanism, in particular the
  appearance of the negative radiation pressure in the $\phi^4$ and $\phi^6$ models.
  We also show that the  process of production of the kink-antikink pairs, induced by radiation is chaotic.
}

%%%%%%%%%%%%%%%%%%%%%%%%%%%%%%%%%%%%%%%%%%%%%%%%%%%%%%%%%%%%%%%%%%%%%%%%%%%%%%%%%%%
%%%%%%%%%%%%%%%%%%%%%%%%%%%%%%%%%%%%%%%%%%%%%%%%%%%%%%%%%%%%%%%%%%%%%%%%%%%%%%%%%%%

%%%%%%%%%%%%%%%%%%%%%%%%%%%%%%%%%%%%%%%%%%%%%%%%%%%%%%%%%%%%%%%%%%%%%%%%%%%%%%%%%%%
%%%%%%%%%%%%%%%%%%%%%%%%%%%%%%%%%%%%%%%%%%%%%%%%%%%%%%%%%%%%%%%%%%%%%%%%%%%%%%%%%%%
\section*{Introduction}
Topological solitons gained increasing interest over the last decades.
Many models which support classical soliton solutions, have been intensively studied
in a wide variety of physical contexts, see e.g. \cite{Vilenkin,Solitons}. Perhaps one of the simplest examples
of solitons is the class of the \textit{kink} configurations, which appears in the (1+1) dimensional
models with a potential possessing  two or more degenerated minima.
The sine-Gordon (sG) model with infinitely degenerated vacuum is a special case of integrable theory. Other models
with polynomial potentials, like for example the simple $\phi^4$ model
with double degenerated vacuum $U(\phi)=\frac12(\phi^2-1)^2$, or the $\phi^6$ model with triple degenerated vacuum, are non-integrable.
The $\phi^4$ model arises in many different physical situations,
it serves as a prototype for many non-linear systems.
%one can say it
%plays the role somewhat similar to the one played by the usual harmonic oscillator in various linear systems.
Indeed, this model is known in the cosmological
context \cite{Vilenkin}, it also  has a number of applications in condensed matter physics
\cite{Solitons}. In particular, it
was applied to describe solitary waves in shape-memory alloys \cite{Falk:1984}, it also can be used as
as a phenomenological theory of the non-perturbative transitions in polyacetylene chain \cite{Rice1979}.
Furthermore, the $\phi^4$  model has been applied in biophysics to describe soliton excitations in DNA double helices
\cite{Yomoza:1983}. The static limit of this model is known as a phenomenological theory of second order phase
transitions \cite{Ginzburg:1950sr}. In quantum field theory it is used as a model example
to investigate transition between perturbative and non-perturbative sectors of the theory \cite{RingEsp,ESPINOSA1990310},
it is also a model of quantum mechanical instanton transitions in double-well potential \cite{Dashen:1974ci}.
Some possible realization was pointed out in buckled graphene ribbons \cite{Yamaletdinov:2017dlz}.

The kink solution is interpolating between two different vacua of the model. This solution is topologically stable.
The most interesting properties of the kinks can be observed in the processes of their scattering and collisions.
Since the models with polynomial potentials are not exactly integrable, one has to encounter the energy loss to radiation in these processes.
Naively, one could expect the collision between a kink and an anti-kink should always lead to annihilation of the
solitons into large amount of radiation. However, the numerical study of the process reveals a far richer pattern
\cite{Anninos:1991un,Belova:1997bq,Campbell:1983xu,Goodman:2005,Makhankov:1978rg,Manton:1996ex,Moshir:1981ja}.
Numerical simulations show that the processes of collisions of a kink and an anti-kink are chaotic,
i.e., for some values of the impact velocity the solitons bounce back, for some other impact velocity,
slightly smaller or larger, they may annihilate via  an intermediate oscillating bion state \cite{Anninos:1991un,Campbell:1983xu}.

More precisely, for initial velocities above the critical value $v_{cr}= 0.2598$ the two incident kinks always
escape to infinity after collision, with  some energy loss due to
radiation. Below $v_{cr}$, the incident waves generically become
trapped, but there is also a complicated pattern of narrow resonance windows, within which the kinks are again
able to escape. This effect is related with reversible energy exchange between the states of perturbative and
non-perturbative sectors of the model.

At the first impact, some part of the kinetic energy of the colliding solitons
is transferred into excitation of the internal modes of the kinks. They then
separate and propagate almost independently, however there is an attractive force
between them. For initial velocities less than the critical value $v_{cr}$, the kinks do not have enough energy
to escape, so they turn back and collide again. At this moment the energy stored in the internal oscillating modes can be
returned to the translational collective mode allowing the kinks to escape, provided that there is a resonance condition between
the time interval between the collisions, and the oscillation period of the internal modes.

Another mechanism is at work in the $\phi^6$ model, where the resonance windows appear due to resonant energy exchange
between the internal oscillating modes trapped by the kink-antikink ($K\bar K$) pair, and the translational mode of the solitons \cite{Dorey:2011yw}.
We also briefly discuss the mechanism where the role of the energy storage is taken over by decaying quasi-normal modes.

The interplay between the states of the perturbative spectrum and
solitons, attracted a lot of attention recently. In particular, it was pointed out that the
excitations of the internal mode of the kink may produce $K\bar K$ pairs \cite{Manton:1996ex}.
Another interesting observation is that  the interaction between the kink
and the scattering modes of the continuous spectrum results in the effect of negative radiative pressure, i.e.,
the $\phi^4$ kink starts to accelerate towards the incoming wave \cite{Forgacs}.

A peculiar feature of many non-linear models, like the $\phi^4$ theory, is that they also
support time dependent non-perturbative solutions which are not captured by the linear analysis.
An interesting example is the breather in the sG model. Since this model is completely integrable,
the states of the continuum are completely separated from the solutions of the field equations.
Thus, the breather has an infinite lifetime because it does not lose its energy into
radiation. However, there are similar quasi-non-dissipative and almost periodic time dependent
configurations in the $\phi^4$ model, the oscillons \cite{Bogolyubsky:1976nx}.
The oscillons appear as quasi-breathers, they are extremely long-lived localised field configurations,
non-harmonically oscillating about the vacuum. Further, the radiation energy losses of an oscillon are
very small, numerical simulations show that in (1+1) dimensional $\phi^4$ theory the oscillon
survives even after millions of oscillations \cite{Fodor:2006zs,Salmi:2012ta}.

The structure of this brief review is as follows.
In the following section we briefly describe properties of the states of the perturbative spectrum of the
$\phi^4$ and $\phi^6$ theories and the corresponding solitons. Then we review the effect of negative
radiative pressure on the kinks in these models. We then discuss correspondence between the oscillons and the internal modes of the
kinks. Production of the kink-antikink pairs from radiation
is briefly discussed in Section \ref{Sec3}. We review the resonance effects in the $K\bar K$ scattering in the $\phi^6$ model
in the Section \ref{Sec4}.
We end with a discussion of the kink boundary scattering in the $\phi^4$ model on a semi-infinite line and the
radiative decay of the boundary mode.

\section{Solitons and perturbations}
\subsection{Spectral structure of small perturbations}
\label{Sec1}
Let us consider the rescaled Lagrangian density of the  $\phi^4$ model with two symmetric vacua $\phi_0=\pm 1$
\begin{equation}  \label{Lag-phi}
L = \frac{1}{2}\left(\partial_t \phi \right)^2-\frac{1}{2}\left( \partial_x \phi \right)^2 -
\frac{1}{2}\left(\phi^2 -1\right)^2.
\end{equation}The corresponding field equation is
\begin{equation}
\label{phi4-eq}
\partial_t^2\phi - \partial_x^2 \phi + 2\phi (\phi^2-1) = 0 \, .
\end{equation}Evidently, there are eigenmodes of the corresponding linearized problem,
which correspond to small oscillations about one of the vacua.
Suppose $\phi(x,t)=1+\xi(x,t)$,
then the expansion in $\xi(x,t)$ yields the linearized equation
\begin{equation}
(\partial_t^2 - \partial_x^2 + 4) \xi(x,t) = 0 \, .
\end{equation}Clearly, this is the usual Klein-Gordon equation for
scalar excitations with mass $m=2$, there is a continuum spectrum of excitations with frequencies
$\omega = \pm \sqrt{k^2+4}$.

The kink configuration is a topologically nontrivial static solution of the equation \re{phi4-eq},
which interpolates between the two vacua, $\phi(-\infty)=\pm 1$, $\phi(\infty)=\mp 1$,
\begin{equation}
\label{phi4-kink}
\phi_K(x) = \tanh(x-x_0)\, ; \qquad \phi_{\bar K} = -\tanh(x-x_0) \, .
\end{equation}Here $\phi_{\bar K}$ is an antikink solution. The kinks are topological solitons,
the field of the kink is a map $\phi: \mathbb{Z}_2 \to \mathbb{Z}_2$.
Physically, they correspond to localized lumps of energy centered around the $x=x_0$.

Let us consider small excitations $\xi(x,t) = \eta(x)e^{i\omega t}$ of the
kink configuration \re{phi4-kink} for $x_0=0$. The corresponding linearized equation is
\begin{equation}
\label{perturb-phi4}
\left(-\frac{d^2}{dx^2}+ V(x)\right)\eta=\omega^2\eta \,,\qquad V(x)=4 - \frac{6}{\cosh^2 x} .
\end{equation}This equation appears as a typical problem in the context of
one-dimensional quantum mechanics, it describes a particle scattering on the  potential $ V(x)$.
Such a potential has two bound modes. The lowest one
\begin{equation}
 \eta_{0}=\frac{1}{\cosh^2 x}\,,\qquad\omega_0=0,
\end{equation}
has an interpretation of a translational mode $\phi_K(x-a)\approx\phi_K(x)-a\eta_0(x)$ and is a reflection of the translational
symmetry of the model $x\to x-a$.
The first mode with odd symmetry,
\begin{equation}
 \eta_{1}=\frac{\tanh x}{\cosh^2 x}\,,\qquad\omega_1=\sqrt{3},
\end{equation}
is called the oscillatory internal mode
\footnote{This mode is also referred to as ``discrete mode" \cite{Manton:1996ex}, or
``wobbling mode" \cite{Segur:1983yv}, or ``shape mode" \cite{CampbellPeyrard,Anninos:1991un}.}.
In the linear approximation this mode oscillates with constant amplitude and frequency.
% This analogy allows us to
% find the spectrum of linear excitations explicitly.
% Let us introduce the ladder operators \cite{Halavanau:2012dv}
% \begin{equation}
% \hat a^\dagger=-\frac{d}{dx}+ 2 \tanh x; \qquad
% \hat a=\frac{d}{dx}+2\tanh x \, ,
% \end{equation}% which form the algebra
% \begin{equation}
% \label{ladder-algebra}
% \left[\hat a, \hat a^\dagger \right] = \frac{4}{\cosh^2 x} \, .
% \end{equation}% It is easy to see that the equation \re{perturb-phi4} then can be written as
% \begin{equation}
% \label{pert-ladder}
% \hat L\eta(x) \equiv \hat a^\dagger \hat a ~\eta(x) = \omega^2 \eta(x) \, .
% \end{equation}% The zero translational mode with eigenvalue $\omega=0$ corresponds to
% the ground state annihilated by the operator $a$, i.e
% \begin{equation}
% \label{zero-mode-eqs-ch2}
% \hat a \eta_0 \equiv \left( \frac{d}{dx}+2 \tanh x\right)\eta(x)= 0 \, .
% \end{equation}% Thus, $\eta_0 = 1/\cosh^2 x$. Notably, the spectrum of fluctuation of
% the $\phi^4$ kink contains an internal oscillating mode, it is produced by the action of the raising operator $\hat a^\dagger$ on the
% zero mode,
% \begin{equation}
% \label{kink-internal}
% \eta_1=\hat a^\dagger \eta_0 = \frac{\sinh x}{\cosh^2 x}\, .
% \end{equation}% The corresponding eigenvalue is $\omega_1=\sqrt 3$.
% Note that the internal mode of the kink does not appear in the perturbative spectrum of the sine-Gordon model.

The continuum modes of the kink with eigenvalues $\omega^2=k^2+4$ are
\begin{equation}
\label{phi4-cont-modes}
\eta_k(x) =\frac{3~\tanh^2 x -3ik~\tanh x - 1 - k^2}{\sqrt{(k^2+1)(k^2+4)}}e^{ikx} \,.
\end{equation}It is worth mentioning that the potential generated by the sG soliton has also the form of the P\"oschl-Teller potential
$V_{sG}(x)=1-2\sech^2x$ and the eigenvalues for $\omega^2=k^2+1$ are also known:
\begin{equation}
\label{sG-cont-modes}
 \eta_k(x) =\frac{ik-\tanh x}{\sqrt{k^2+1}}e^{ikx} \,.
\end{equation}
Note that in  the sG model the kink has no internal oscillating mode, and the translational mode $\eta_0=\sech x$ is the only bound state of the
linearized potential.

An interesting generalization of the double vacuum model \re{Lag-phi} is the 1+1 dimensional $\phi^6$ model, which
is defined by the Lagrangian  \cite{Lohe:1979mh}
\begin{equation}  \label{Lag-phi6}
L = \frac{1}{2}\left(\partial_t \phi\right)^2-\frac{1}{2}\left(\partial_x \phi\right)^2 - \frac{1}{2}\phi^2\left(\phi^2 -1\right)^2.
\end{equation}Clearly, there are three degenerated vacua,
$\phi_0 \in \left\{-1,0,1\right\}$, however, like $\phi^4$ theory, the model \re{Lag-phi6}
is symmetric with respect to reflectional $\mathbb{Z}_2$ symmetry
$\phi \to -\phi$ and/or $x\to -x$. The resulting equation of motion
\begin{equation}
\label{phi6-eq}
\partial_t^2\phi - \partial_x^2 \phi + \phi - 4 \phi^3 + 3 \phi^5 = 0 \, ,
\end{equation}can be linearized about each of three vacua,
the excitations $\xi(x,t)$ satisfy the linearized equations
\begin{equation}
\label{pert-pot-phi6-triv}
\begin{split}
&(\partial_t^2 - \partial_x^2 + 1) \xi(x,t) = 0; \quad {\rm as }~~ \phi_0= 0\, ;\\
&(\partial_t^2 - \partial_x^2 + 4) \xi(x,t) = 0; \quad {\rm as }~~ \phi_0=\pm 1 \, .
\end{split}
\end{equation}Thus, the scalar excitations about the vacua $\phi_0=\pm 1$ have mass $m_1=2$ while the excitations about the symmetric
vacuum $\phi_0=\pm 0$ have mass $m_2 =1$.

The $\phi^6$ model is quite richer in terms of its solitonic solution than the $\phi^4$ theory.
Indeed, there are two different kinks interpolating between neighboring
vacua $\phi(-\infty)=0$, $\phi(\infty)= 1$ and $\phi(-\infty)=- 1$, $\phi(\infty)= 0$:
\begin{equation}
 \label{kink-phi6}
 \phi_{(0,\pm 1)}(x) = \pm\sqrt{\frac{1 + \tanh(x)}{2}}\,,\qquad\phi_{(\pm 1,0)}(x)=\phi_{(0, \pm 1)}(-x).
\end{equation}

% \begin{equation}
% \label{kink-phi6}
% \begin{split}
% &\phi_{(0,1)} = \sqrt{\frac{1 + \tanh(x-x_0)}{2}} = \sqrt{\frac{1}{1+e^{-2(x-x_0)}}}; \\
% &\phi_{(-1,0)} = -\sqrt{\frac{1 - \tanh(x-x_0)}{2}} = -\sqrt{\frac{1}{1+e^{2(x-x_0)}}} \, .
% \end{split}
% \end{equation}% Evidently, there are translational zero modes \index{Zero mode} in the spectrum of excitations of the kinks,
% \begin{equation}
% \label{phi6-kink-zero}
% \eta_{1}^{(0)} = e^{2x}\left(\frac{1}{1+e^{2x}} \right)^{3/2};\qquad
% \eta_{2}^{(0)} = e^{-2x}\left(\frac{1}{1+e^{-2x}} \right)^{3/2} \, .
% \end{equation}
For an isolated $\phi^6$ kink, however, there are no localized bound state solutions to the
linearized $\phi^6$ equations with the potential $V(x)=15 \phi_K^4-12 \phi_K^2+1$.
General continuum solutions of this equation can be written in terms of hypergeometric functions \cite{Lohe:1979mh}.
A special feature of the spectrum of linear perturbation around the $\phi^6$ kinks
is that, unlike the $\phi^4$ model, the potential of the linearized problem $V$
is not symmetric with respect to reflections $x \to -x$. In other words,
the mass of the excitations of the continuum is different on the opposite sides of the kink \cite{Dorey:2011yw}.

\section{Interplay between the states of perturbative and non-perturbative sectors}
\label{Sec2}
\subsection{The effect of negative radiation pressure}
Solitons in many ways are very similar to particles, yet there are some important differences.
The internal structure of solitons may be revealed, for example, in the scattering spectrum of
an external incoming perturbations. On the other hand, the scattering on the soliton results in the
radiation pressure exerted on it, this effect is similar to the
familiar solar radiation pressure force on the particles in comets' tails.

Let us consider a small amplitude propagating perturbation of the scalar field, which represents a
wave moving towards a soliton.
In general, we can expect that the scattering spectrum contains both the
reflected and transmitted waves, thus
\begin{equation}
 \phi({x\to-\infty})=\frac{1}{2}A\left[e^{i(\omega t-kx)}+Re^{i(\omega t+k x)}\right]+c.c \, .
\end{equation}
and
\begin{equation}
 \phi({x\to+\infty})=\frac{1}{2}ATe^{i(\omega t-k x)}+c.c \, ,
\end{equation}
where $A$ is the amplitude of the incoming wave, $\omega$ and $k$ are the frequency and the wave number of the wave
and $R$, $T$ are reflection and transmission coefficients, respectively.

The wave carries both the energy
\begin{equation}
\mathcal{E} = \frac12\left(\phi_t^2 + \phi_x^2\right) + U(\phi)
\end{equation}and the  momentum density
\begin{equation}
\mathcal{P}=\phi_x\phi_t\, .
\end{equation}
The corresponding continuity equation is
\begin{equation}
 \partial_t\mathcal{P}=-\frac{1}{2}\partial_x\left(\phi_t^2+\phi_x^2-2U(\phi)\right)\, ,
 \label{flux}
\end{equation}
thus, there is a force acting on the soliton due to the transfer of
momentum from the incoming radiation.

We can now evaluate this force
averaging the flux \re{flux} over the period and integrating it by parts.  We obtain
\begin{equation}
 F= \overline{\partial_t\mathcal{P}} = \frac{1}{2}k^2A^2\left(1+|R|^2-|T|^2\right) \ .
\end{equation}
Further, in the case of a single channel scattering we can make use of the
continuity equation $|R|^2+|T|^2=1$. In such a case this formula simplifies to
$F=A^2k^2|R|^2$. Note that in the single channel scattering the radiative force always pushes the soliton
in the direction of the incoming momentum.

One of the most surprising features of both sG and $\phi^4$ model is that the solitons do not
reflect any radiation in the linear
approximation. The reflection coefficients read from the solutions of the linearized equations
(\ref{phi4-cont-modes}) and (\ref{sG-cont-modes}) are
exactly 0, there is no part proportional to $e^{-ikx}$. This is a general feature of the
P\"oschl-Teller potential $V(x)=-\frac{n(n+1)}{\cosh^2(x)}$ with  $n\in\mathbb{Z}$.

The question arises how exactly the soliton would move when it is exposed to the radiation in
the full non-linearized system. The sG model is integrable, there are analytical solutions
corresponding to a static soliton with a moving cnoidal
wave in its background \cite{Shin,Forgacs}. However, the sG soliton is completely transparent to the wave in all orders.
The interaction is not dynamical, there is no
energy nor momentum transfer between the incoming wave and the soliton, the interaction just results in a phase shift.
The $\phi^4$ model, on the other hand, is not integrable, some interesting effects arise there already
in the second order of the perturbation series.

We search the solution of the dynamical equation in the form of a perturbation series with amplitude of
incoming wave  $A$ as an expansion parameter:
\begin{equation}
\phi=\phi_s+\xi=\phi_s+A\xi^{(1)}+A^2\xi^{(2)}+\cdots\, .
\label{eq:post1}
\end{equation}
The equation in the $n^\textrm{th}$ order has a general form
\begin{equation}\label{eq:gen0}
\ddot\xi^{(n)}+\opL\xi^{(n)}=f^{(n)} \, ,
\end{equation}
where the linear operator is $\opL=-d^2/dx^2+U''(\phi_s(x))$, and the source term $f^{(n)}$ depends on
the solutions of lower
orders. In particular $f^{(1)}=0$. By taking the monochromatic wave
$\xi^{(1)}=\frac{1}{2}e^{i\omega t}\eta_{-q}(x)+c.c.$, where
$q=\sqrt{\omega^2-4}>0$ is the wave number, we can evaluate the source term in the second order:
\begin{equation}
f^{(2)} = -6\phi_s{\xi^{(1)}}^2=-\frac{3}{2}\left(e^{2i\omega t}\eta_{-q}^2+2\eta_q\eta_{-q}+e^{-2i\omega t}\eta_{q}^2\right)\,.
\end{equation}
The solution can be sought in the form
\begin{equation}
 \xi^{(2)}=\xi^{(2)}_{+2}e^{2i\omega t}+\xi^{(2)}_{0}+\xi^{(2)}_{-2}e^{-2i\omega t}\,.
\end{equation}
Similarly, in general in the $m^\text{th}$ order the solution can be expanded in a Fourier series of $\xi^{(m)}_n$
functions oscillating with the
frequency $n\omega$ ($|n|\leq m$). All these solutions have an asymptotic form consisting of the
inhomogeneous part, which originates from the source term, and outgoing wave:
\begin{equation}
 \xi^{(m)}_n(x\to\pm \infty)=\eta_{inh}+\alpha_{mn,\pm k}\eta_{\mp k},\,\qquad k=\sqrt{n^2\omega^2-4}\,.
\end{equation}
The coefficients  $\alpha_{mn,\pm k}$ can be treated as nonlinear scattering amplitudes, they can be found
using the Green's function technique:
\begin{equation}
 \alpha_{mn,\pm k}=-\frac{1}{W}\int_{-\infty}^\infty dx'\,\eta_k(x')f^{(m)}_n(x')\,,
\end{equation}
where $W=\eta_k'\eta_{-k}-\eta_k\eta_{-k}'$ is the Wronskian. Knowing those amplitudes we can calculate the
momentum balance on both sides of the
kink and hence find the force which is exerted by the wave.

Since, in the $\phi^4$ the kinks are transparent in the first order, the first non-vanishing
contribution to the force is proportional to the \textit{fourth} power of the amplitude:
\begin{equation}
 F^{(4)}=2A^4k\left[(k+2q)|\alpha^2_{22,k}|-(k-2q)|\alpha^2_{22,-k}|\right]\,,
\end{equation}
where $q=\sqrt{\omega^2-4}$ and $k=\sqrt{4\omega^2-4}$ are the wave numbers for the
frequencies $\omega$ and $2\omega$ respectively.
The coefficients can be found in a closed form
\begin{equation}\label{alpha22-fi4}
\alpha_{22, k}(q)
= -\frac{3}{2}\pi\frac{q^2+4}{q^2+1}\sqrt{\frac{q^2+4}{k^2+1}}\;
\frac{1}{k\sinh\left(\frac{2q+k}{2}\pi\right)}\,,
\end{equation}
thus, the force is
\begin{equation}\label{force}
F^{(4)}=A^{4}f^{(4)}=\frac{9\pi^2A^4\omega^6}{k(4\omega^2-3)(\omega^2-3)^2}\left[ \frac{\omega_+}{\sinh^2\pi\omega_+}-
\frac{\omega_-}{\sinh^2\pi\omega_-}\right] \, ,
\end{equation}
where
\begin{equation}
 \omega_\pm:=\sqrt{\omega^2-1}\pm \sqrt{\omega^2-4}\,.
\end{equation}
Note that the force is negative,  although the incoming wave
is traveling from $-\infty$, the kink accelerates towards negative values
of $x$. In other words, the kink is pulled by the \textit{negative radiation pressure} (NRP).
This surprising result was confirmed by numerical simulations of the full nonlinear partial differential equation.
Both the $A^4$ proportionality and the frequency dependence of the force \re{force} was confirmed for the
frequency range $\omega<5$ and $A<0.22$ with 10\% accuracy. The only
discrepancy near the frequency $2\omega_d$ could be explained by the highly nonlinear resonance
with the internal oscillating mode of the kink.

The effect of NRP can be surprising and counterintuitive at first. However,
the physical explanation of this effect is relatively simple. In the first
order the kink is transparent to the incoming radiation,  the force exerted on the kink is zero.
Nonlinearities around the kink, on the other hand, produce higher frequency waves which also carry momentum.
Since the second order transmission coefficient is
larger than the corresponding reflection coefficient, there a surplus of the momentum behind the kink.
In order to comply with the conservation low and restore the balance
the kink starts to accelerate in opposite direction, towards the source of radiation.

The negative radiation pressure in the $\phi^4$ model seems to be something exceptional. It
requires that the potential, generated by the soliton configuration in a non-integrable model,
is reflectionless. This is rather rare property.
However, one can expect that the effect of NRP still may exist in some modifications of the $\phi^4$ model with
non-reflectionless perturbation potential. In such a case  the
force exerted on the soliton would have the form
\begin{equation}
 F\approx A^2f^{(2)}-A^4f^{(4)}+\mathcal{O}(A^6)\,.
\end{equation}
This would support the effect of NRP  for certain range of amplitudes, as \mbox{$f^{(2)}<A^2f^{(4)}$}.

The mechanism of the NRP we described above is not unique. Indeed,
the key point is to produce a surplus of momentum carried by the radiation behind the kink.
This can be achieved in many different ways.
For example, in more complicated multicomponent field theories,
different components can have different dynamical properties.
Then, even in the linear order, the scattering with increase of momentum becomes allowed.
For example a scattering with transfer of momentum from the massive field $\mathcal{P}_t\sim (\omega^2-m^2)$ to
the massless field $\mathcal{P}_t\sim \omega^2$ leads to the
NRP, as long as the reflection from the scattering center remains small enough \cite{Romanczukiewicz:2008hi, Forgacs:2013oda}.

Quite interesting, and even simpler example of NRP effect, can be found in the $\phi^6$ model \re{Lag-phi6}
\cite{trom2017nrp1}.
In this model we restrict our considerations only the the first, linear order.
Recall that the model \re{Lag-phi6} has three vacua.
Small perturbations around two of them possess mass $m_{\pm 1}=2$, however, the perturbations
around the third vacuum have smaller mass  $m_0=1$.
The kink interpolates between vacua of different kinds, so is a sort of a bridge for the waves
which can travel from one vacuum to another.
Waves with the frequency $m_{0}<\omega<m_{\pm1}$ can only propagate around the vacuum $\phi=0$ and
reflect perfectly from the kink.
This gives raise to a \textit{positive radiation pressure}.
But also for higher frequencies, even when they can already propagate in the second vacuum  they move
slower and carry less momentum.
This deficiency of momentum is balanced by the force pushing the kink towards
$\phi=\pm1$ vacuum.

On the contrary, the wave traveling initially through $\phi=\pm 1$ vacuum carries less momentum than after
the transition through the kink. Behind
the kink the excess of momentum is created pushing the kink again towards the $\phi=\pm1$ vacuum.
Such a wave exerts negative radiation pressure. Indeed, the solutions of the linearized
equation for  perturbations around the $\phi^6$ kink are known in terms of hypergeometric functions \cite{Lohe:1979mh}, thus
the scattering amplitudes can be found from the asymptotic
form of the solution
\begin{equation}
   \begin{cases}
      \eta(x\to+\infty)=e^{ikx}/B(q,k),\\
      \eta(x\to-\infty)= e^{iqx}+\frac{B(-q, k)}{B(q,k)}e^{-iqx}
   \end{cases}
\end{equation}
with
\begin{equation}
   \begin{gathered}
   q=\sqrt{\omega^2-1},\;k=\sqrt{\omega^2-4},\\
   B(q,k) = \frac{\Gamma(1-ik)\Gamma(-iq)}{\Gamma(-\frac{1}{2}ik-\frac{1}{2}iq+\frac{5}{2})\Gamma(-\frac{1}{2}ik-\frac{1}{2}iq-\frac{3}{2})}.
   \end{gathered}
\end{equation}
Then the force exerted on the $\phi_{(0,1)}$ kink by the
incoming wave, propagating from  $+\infty$, can be evaluated as
\begin{equation}
   F_{+\infty}(q,k) = \frac{1}{2}\frac{A^2}{|B(q,k)|^2}\left(2|B(-q,k)|^2q^2+qk-k^2\right)\,.
\end{equation}
Similarly the force acting on the $\phi_{(0,1)}$ kink by a wave propagating in opposite direction, can be expressed as
$F_{-\infty}(q,k) = -F_{+\infty}(k,q)$.
Both forces are positive, which means that no matter from which direction the wave came, the kink would always
accelerate rightwards.
Moreover, because the force appears in the linearized system, in the general case of an arbitrary perturbation
the contribution of all modes can be evaluated as an additive sum over all eigenfrequencies.
Therefore we can say that \textit{arbitrary} small perturbations will always push the  $\phi_{(0,1)}$ kink towards the
vacuum $\phi=1$.
Considering all kinks of the $\phi^6$ model we can conclude that all small perturbations will act on the solitons
in such a way that the vacuum $\phi=0$ will tend to
expand.

\begin{figure}
 \begin{center}
 \includegraphics[width=1\linewidth]{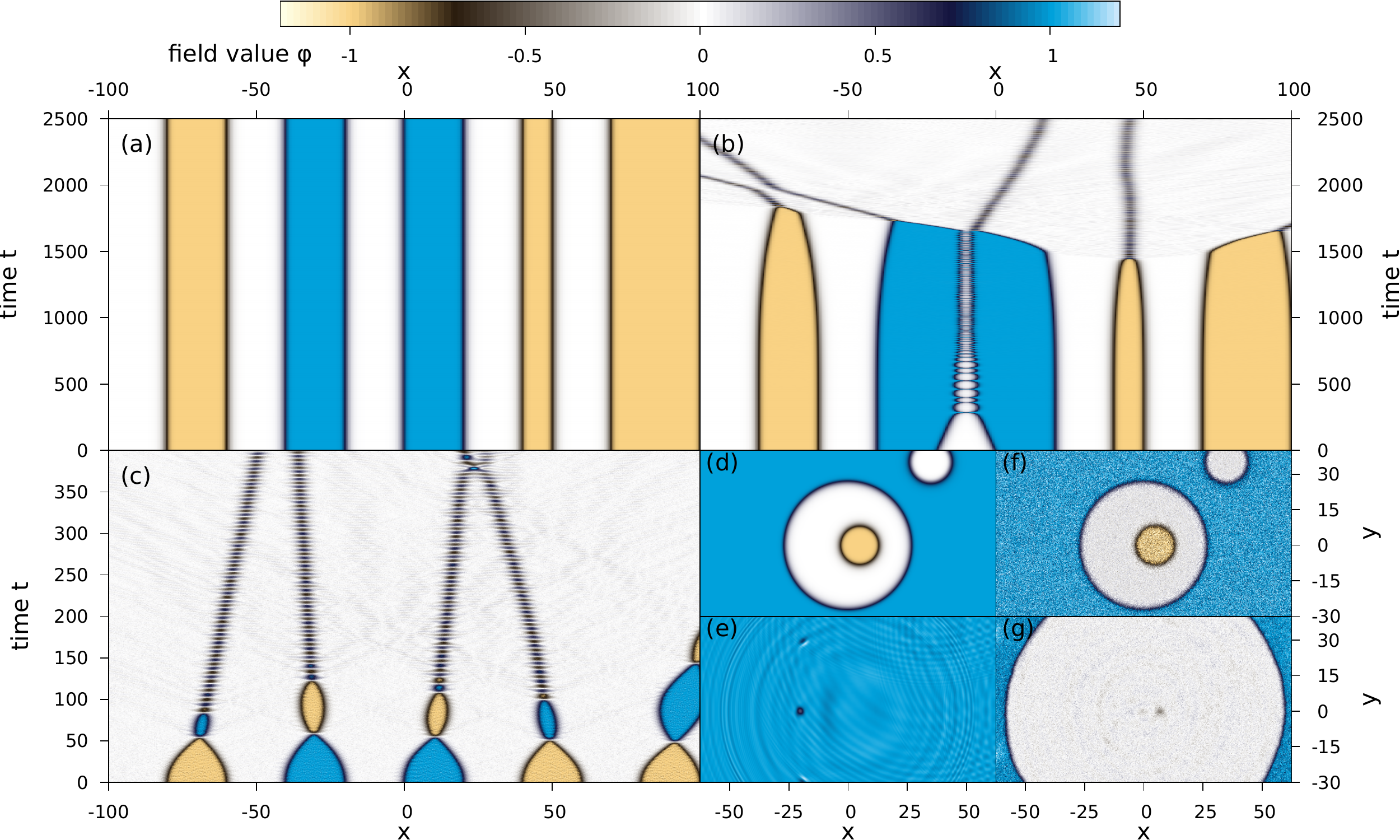}
 \caption{\small Evolution of system of kinks in $\phi^6$ model: (a) nearly static configuration,
 (b) annihilation of close kink-antikink pair triggers the chain reaction of other annihilation,
 (c) with gaussian
noise with amplitude $A=0.05$. Evolution of two dimensional circular domain walls: without perturbation (d) $t=0$ (e) $t=75$ and with gaussian
noise $\langle A\rangle=0.12$, (f) $t=0$ (g) $t=75$.
Reprinted (without modification) \copyright2017 The Author of \cite{Romanczukiewicz:2017hdu}, Published by Elsevier B.V.}\label{fig:phi6Evolutions}
\end{center}
\end{figure}

This is a highly nontrivial conclusion with many important implications.
Let us consider a system of (almost) stationary kinks and anti-kinks separated by a distance so large that the
interactions are almost negligible.
The system would last in such a state for a very long time, given the fact that the static forces decay exponentially.
However, it is enough to put any radiative perturbation, which can be either a localized
perturbation or even random fluctuations filling whole space
(just like the primordial, background radiation filling the whole Universe) and the whole system would collapse.
Moreover, the system would evolve in such a way, that the defects enclosing one of the vacua $\pm 1$ would collide.
Such collisions are the source of radiation which would increase the rate of further collisions.
This is how an interplay between negative and positive radiation pressure can lead to a chain reaction
of kink-antikink annihilations.
This scenario was confirmed by numerical simulations (Figure \ref{fig:phi6Evolutions} a,b,c).

In higher ($D$) dimensions, instead of kinks there are $D-1$ dimensional domain walls.
The walls can extend to either infinities or enclose domains of certain vacua.
Large domain walls can be locally flat and the dynamics can be very similar to the dynamics of kinks.
However, curved domain walls can have a surface tension which tries to close small domains \cite{ArodzPelka2000} (Figure \ref{fig:phi6Evolutions}
(d)$\to$(e)).
The radiation pressure can increase the rate in which domains with $\phi=\pm 1$ vacua
vanish and slow down the rate or even reverse for the $\phi=0$ vacuum.
For instance, random fluctuations can expand circular or spherical domains of the $\phi=0$ vacuum above certain, critical radius (Figure
\ref{fig:phi6Evolutions} (f)$\to$(g)).

\subsection{From internal modes to oscillons and back}
Apart from kinks, the simple $\phi^4$ model \re{Lag-phi} also possess a very interesting regular quasi-non-dissipative and almost periodic
time-dependent solution,
whose properties closely resemble the sG breather \cite{Kudryavtsev:1975dj}.
Such a state, observed in the process of time evolution of some initial data, is
referred to as \footnote{In the 3+1 dimensional theory the corresponding spherically symmetric solutions were originally
discovered in 1976 by Bogolyubskii and Makhan'kov \cite{Bogolyubsky:1976nx}, who coined a term ``pulson" to describe these configurations. However
this observation did not attract much attention at that time, this work was almost forgotten until Gleiser rediscovered these solutions in 1994
\cite{Copeland:1995fq,Gleiser}.}
an oscillon.

Since the life time of the oscillons is very large they can be well approximated by the Fourier decomposition \cite{Kosevich}
\begin{equation}
 \phi(x,t) = 1+\sum_{n=0}^N\phi_n(x)\cos(n\omega t)\, .
\end{equation}
During the evolution the amplitude of the oscillon slowly decreases, however, the rate in which it
radiates decreases even faster, the decay rate is
$ \frac{dE}{dt}\sim\exp(-B/E) $ beyond all orders \cite{Boyd,Kruskal-Segur,Fodor:2008du}. The decreases in the amplitude
are synchronized with the increases of the frequency of the oscillations.
In the long time evolution the oscillon tends to the lowest mode of the continuous spectrum with frequency $\omega=2$.

There is an interesting correspondence between the internal modes
and the oscillon, which under special conditions, can smoothly be transferred into each other \cite{Romanczukiewicz:2017gxb}.
Let us consider the modification of the Lagrangian \re{Lag-phi}
\begin{equation}
 \mathcal{L}=\frac{1}{2}\left(\partial_t \phi\right)^2-\frac{1}{2}\left(\partial_x \phi\right)^2
 -\frac{1}{2}\left(\phi^2-1\right)^2-\frac{1}{2}V(x)(\phi-1)^2 \, ,
 \label{lag}
\end{equation}
where an additional perturbation is produced by the asymptotically vanishing
P\"oschl-Teller potential $V(x)=-V_0\sech^2 bx$. Unlike the original $\phi^4$ theory \re{Lag-phi}
the model \re{lag} has only one trivial vacuum $\phi_0=1$. In the second topological sector the static configurations,
which asymptotically tend to $\phi\to -1$ as $x \to \pm \infty$, are nontrivial solutions localized by the trapping
potential $V(x)$. One type of solutions is similar to the non-topological soliton (lump), which appears in
the two-component system of coupled fields \cite{Rajaraman:1978kd,Halavanau:2012dv}, however, the lump is captured by the
potential, it does not propagate.

Solutions of the second type in this sector represent a static kink-antikink pair, trapped by the potential $V(x)$.
This solution is unstable and a small perturbation
can destabilize the configuration which can either decay into the vacuum $\phi_0=1$ with two kinks escaping to infinity,
or into the lump, which is a minimal energy solution in  the same sector. Evidently, a particular choice of
scenario depends on the explicit value of the parameter $V_0$ \cite{Romanczukiewicz:2017gxb}.

The perturbative spectrum of linearized fluctuations around the trivial vacuum $\phi_0=1$ of the modified model
\re{lag} can be found by analogy with consideration above. The corresponding eigenfunctions are
solutions of the equation, which generalizes the linearized equation of the $\phi^4$ model \re{perturb-phi4}
\begin{equation}
\label{perturb}
\left(-\frac{d^2}{dx^2}+ 4 -\frac{V_0}{\cosh^2 bx} \right)\eta=\omega^2\eta \, .
\end{equation}with the mass threshold $m=2$. The ground state of the P\"oschl-Teller potential is
\begin{equation}
\label{eq:linsol}
 \xi_0(x) = \frac{1}{\cosh^{\lambda}(bx)}, \, \, \,  \lambda(\lambda+1)=\frac{V_0}{b^2},\;\,\, \, \omega^2=m^2-b^2\lambda^2
\end{equation}
where $\lambda$ is a  real parameter.

The eigenvalues $\omega$ are imaginary for $V_0>4+2b$, the corresponding modes are unstable and the system
could change its ground state producing a $K\bar K$ pair. There is a single stable oscillating mode for $2>V_0>0$,
as this mode is excited to a non-linear regime, it decays via radiation propagating through the second
harmonic. The amplitude of this mode decreases according to the Manton-Merabet power law $A~t^{-1/2}$
\cite{Manton:1996ex}, its frequency  increases with time, as the mode evolves toward the linear regime.

Numerical analysis reveals that in the presence of the perturbation potential,
the lowest bounded oscillating mode could become an effective  attractor in the time evolution of
some even initial data, which, in the limit $V_0 = 0$ would evolve into the oscillon-like state.

Considering the time evolution of the symmetric initial data
$\phi(x,0)=1-A_0\sech^\alpha(x), ~\phi_t(x,0)=0$, where $\alpha $ is a positive
real parameter, we observe that the absence of the potential, or in case of the repulsive potential,
these initial data evolve into the oscillon solution.

\begin{figure}[!h]
 \begin{center}
\includegraphics[width=0.75\columnwidth]{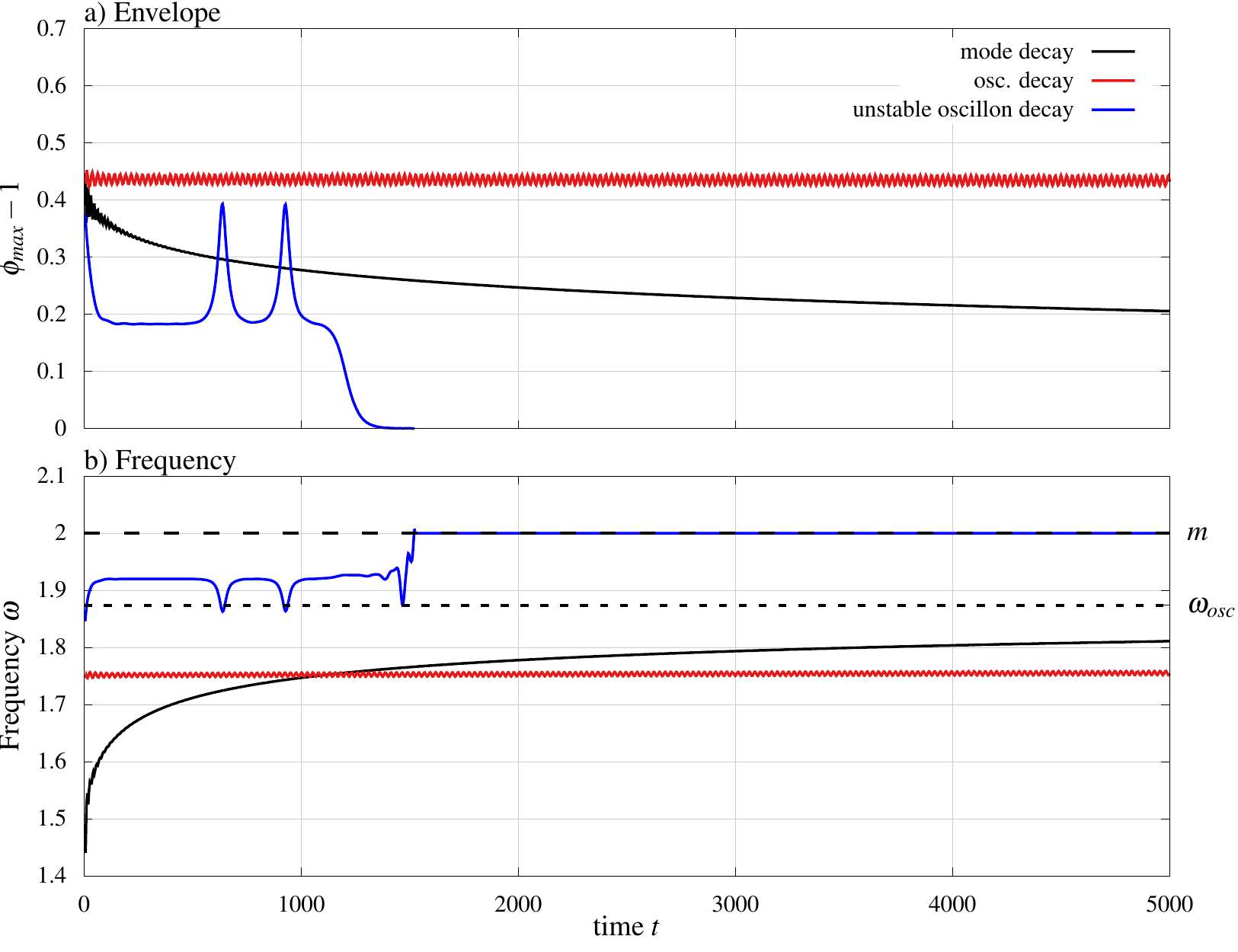}
%\vspace**{-8mm}
\caption{\small  Possible relaxation scenarios for $A_0=0.6$. The plots show
the maximum envelope of oscillations at $x=0$
(upper plot) and the measured frequency $\omega$. Black curve represents
decay of the oscillating mode, red curve shows the time evolution of the oscillon.
Blue curve shows decay of the unstable oscillon in the repulsive potential $V_0<0$.
Dashed horizontal line indicates the eigenfrequency of the mode for $V_0=1.19$, $\omega_{osc}=1.8735$.
Reprinted (without modification) from \cite{Romanczukiewicz:2017gxb}, \copyright
2017 The Authors of \cite{Romanczukiewicz:2017gxb}, under
the CC BY 4.0 license.}
\label{fig:relaxationLong1}
%\vspace**{8mm}
\end{center}
\end{figure}

The red curve in  Fig.~\ref{fig:relaxationLong1} illustrates the time evolution of the oscillon,
which is produced in the absence of the external potential. The black curve corresponds to the evolution of
the oscillating mode, which was excited up to nonlinear regime.
Due to the nonlinearity, the corresponding initial frequency is much lower initially than its value
predicted from the linearized theory.
However with time the energy is radiated away and the amplitude of oscillations slowly decays.
Then the frequency becomes larger approaching the corresponding linearized value.

\begin{figure}[!h]
 \begin{center}
\includegraphics[width=0.75\linewidth]{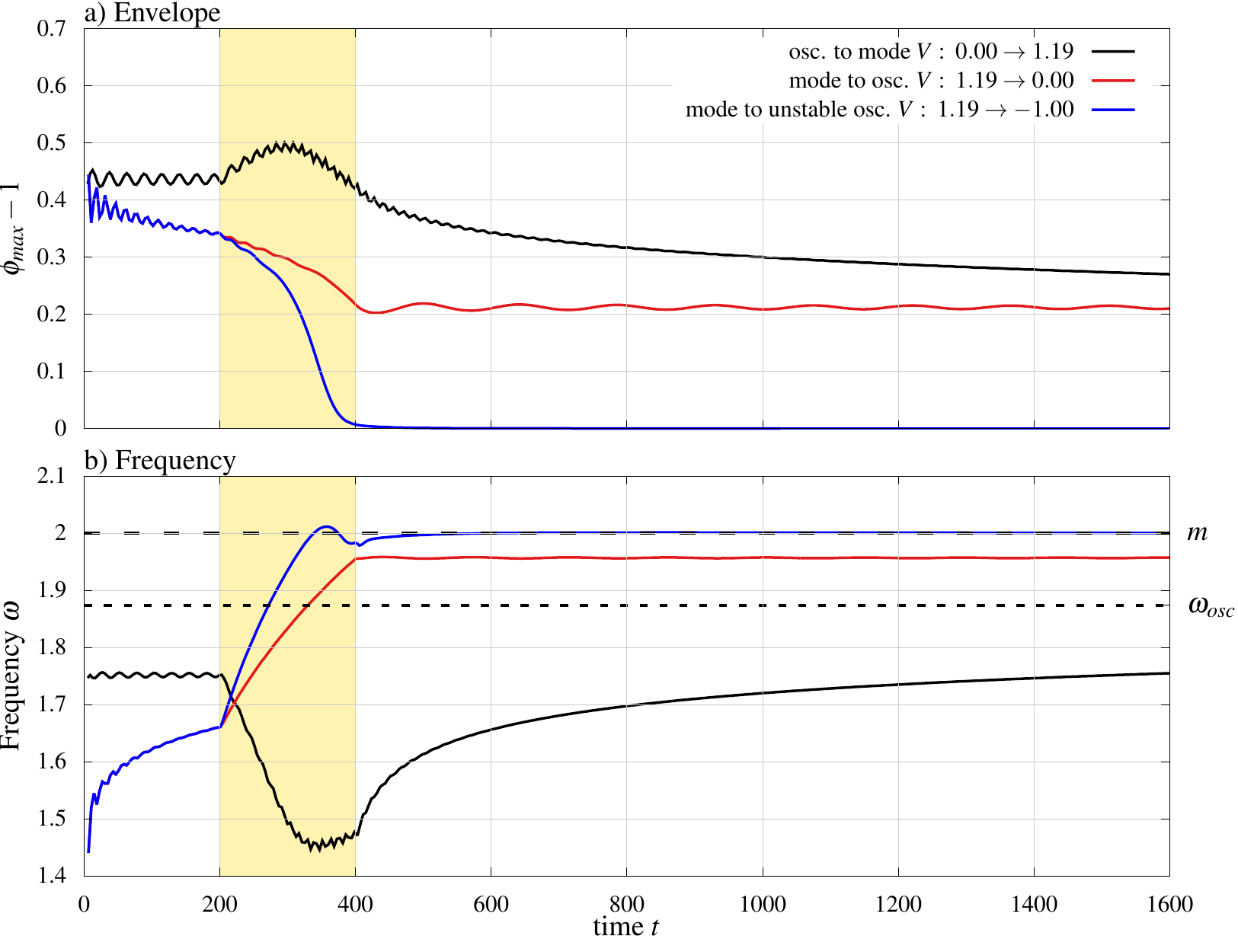}
%\vspace**{-8mm}
\caption{\small  Examples of adiabatic transformations: (i)
the oscillon $A_0=0.6$ to the internal oscillating mode  (black curve);
(ii) the oscillating mode to the oscillon (red curve);
(iii) the oscillating mode to an unstable oscillon and its consequent decay (blue curve).
Reprinted (without modification) from \cite{Romanczukiewicz:2017gxb}, \copyright
2017 The Authors of \cite{Romanczukiewicz:2017gxb}, under
the CC BY 4.0 license.}
\label{fig:relaxationLong}
%\vspace**{8mm}
\end{center}
\end{figure}

An interesting scenario was observed for the situation when the initial data set
rapidly converges to the oscillon configuration in the absence of the external potential, see black curves in
Fig.~\ref{fig:relaxationLong}.
Within the time interval $t \in [t_1, t_2]$,
we adiabatically increase the depth of the potential as
$ V(t) =V_0\frac{t-t_1}{t_2-t_1} $. Then the amplitude of the oscillations increases and,
as  $ t > t_2$, the oscillating state becomes trapped by the potential well.
Consequently, the pattern of evolution of this mode follows the usual
$t^{-1/2}$ law of the radiative decay.
In other words, the oscillon state becomes smoothly transformed into the oscillating mode.

The opposite transition is also illustrated in Fig.~\ref{fig:relaxationLong} (red curves).
Initially, there is an
excited oscillating mode trapped in the potential well with the same values of the parameters
as above. As $t_1<t<t_2$ the potential is
turned off smoothly
and the oscillation mode becomes transformed into the oscillon state with
amplitude  a bit above $A=0.2$ and the frequency $\omega=1.9566$. This is clearly below the mass
threshold. Both the frequency and the amplitude of the configuration are modulated by small oscillations,
however the corresponding average values are almost constant. Thus,
we can conclude that a distinction between an oscillon and an oscillating mode is quite artificial.
The only difference is that the frequency of the oscillating modes tend to the frequency,
calculated from the linearized model, as the amplitude decreases.

\section{Production of the kink-antikink pairs from radiation}
\label{Sec3}
As we have seen, there is  strong evidence of
intrinsic relations between the states of the perturbative spectrum of a nonlinear
non-integrable theory and the non-perturbative soliton solutions.
This interplay significantly affects the process of collision of the solitons in the $\phi^4$ model,
there is an intriguing pattern of scattering of the kinks
\cite{Anninos:1991un,Belova:1997bq,Campbell:1983xu,Goodman:2005} related
with resonant energy exchange between the translational mode of the solitons
and excitation of the internal vibrational mode. Moreover, the
opposite process of the production of the $K\bar K$
pairs in the collision of the states of continuum also is chaotic due to resonance effects \cite{Romanczukiewicz:2005rm, Romanczukiewicz:2010eg}.

We considered two widely separated  wave trains with amplitude $C$
propagating towards the collision point which is a kink or one of the vacuum sectors of the $\phi^4$ model.
The initial data are
\begin{equation}
\label{initial}
   \phi(x,t) = \phi_0 + C [F(x+vt)\sin(\omega t+kx) +F(x-vt)\sin(\omega t-kx)],
\end{equation}where $k$ is the wavenumber of the incoming wave, $\omega=\sqrt{k^2+4}$
is the frequency of the continuum mode and $v=k/\omega$ is the velocity of the propagation of the
wave train. The envelope of the train can be taken as $F(x) =[\tanh (x-a_1)-\tanh(x-a_2)]$, here the
parameters $a_1$ and $a_2$ define the length of the train and initial separation between the trains.
$\phi_0$ is the static solution, which in the case of $\phi^4$ can be either a vacuum or a single kink.

In the case of a kink $\phi_0=\phi_K$ the small amplitude radiation interacts with the internal mode of the kink.
The mode is excited due to a nonlinear parametric resonance (similar to the Mathieu equation) \cite{Romanczukiewicz:2005rm}.
As the amplitude of the radiation grows, the response of the mode also grows.
For certain amplitude the internal oscillating mode is excited so much that the energy can be released only by ejecting a $K\bar K$ pair.
A scan through both the amplitude and frequency revealed that the creation process is chaotic, and the boundary between creation and just the
excitation of the internal mode has some fractal properties.
The absorption of the radiation by the oscillating mode is responsible for the aforementioned discrepancy between the second order calculations of
the negative radiation pressure and the numerical results.

In the absence of the kink, i.e., for $\phi_0=1$, the  numerical simulations show that the small amplitude collision produces an oscillating lump
with the frequency just a bit above the mass threshold. This lump could be identified with a non-linear  excitation of the trivial vacuum, it slowly
radiates its energy away.
For large amplitude collisions, the remaining lump oscillates with frequency within the mass gap, so such a state can be identified as an oscillon.

\begin{figure}
   \centering
   \includegraphics[width=0.75\columnwidth]{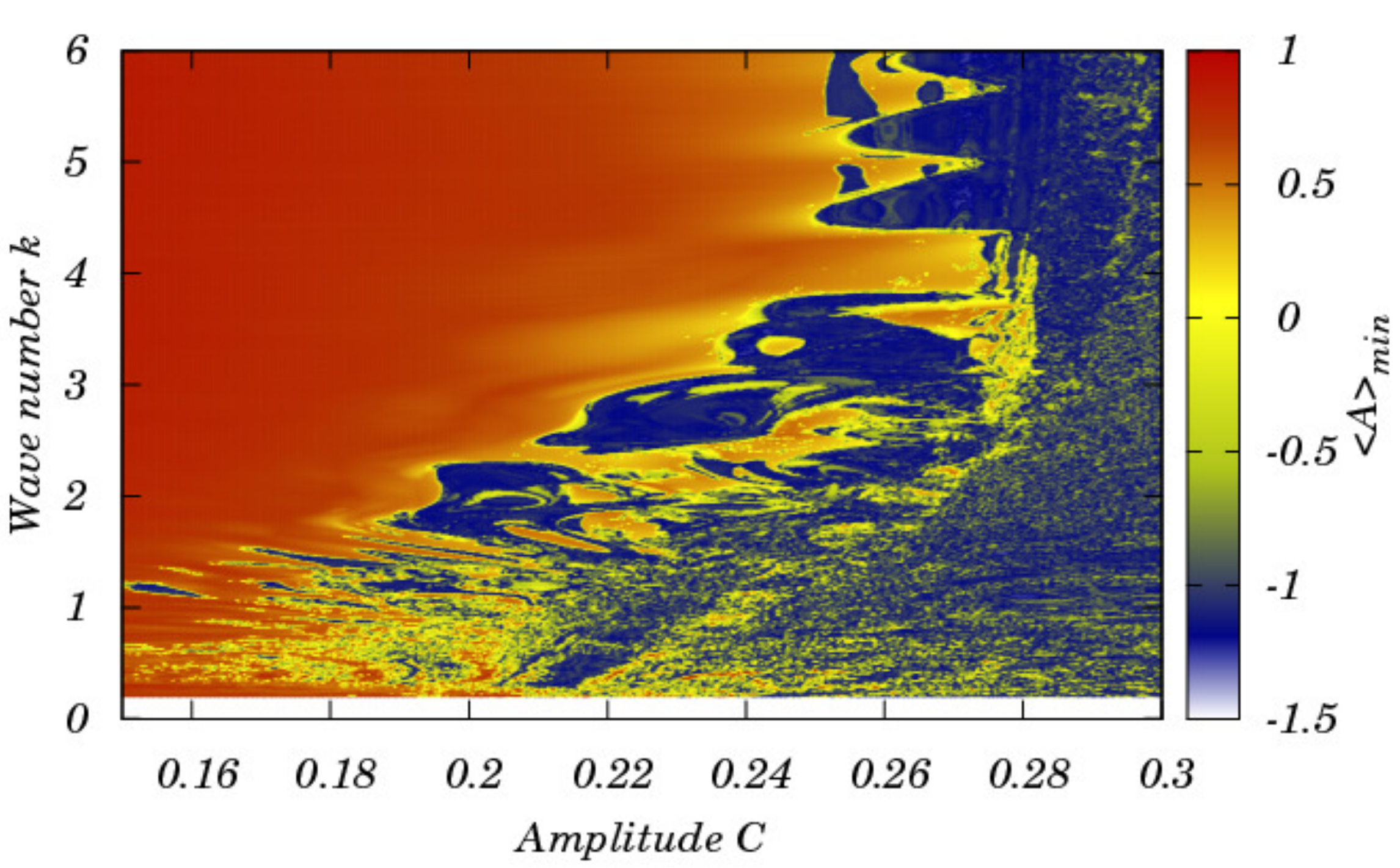}
   \caption{\label{fig:Fractal2}\small Fractal structure in the $C,k$ plane. Shading (or colour)
represent the measured minimum of average of the field
$\langle A\rangle=\frac{1}{20}\int_{-10}^{10}\!dx\; \phi(x,t)$.
The dark regions (blue in colour), where  $\langle A\rangle<-1$ indicate creation of the $K\tilde K$ pairs.
Reprinted (without modification) from \cite{Romanczukiewicz:2010eg}, \copyright
2010 APS, the Authors of \cite{Romanczukiewicz:2010eg}, under
the copyright license.}
\end{figure}

Furthermore, for a certain range of impact parameter values,
the $K\bar{K}$ pair produced in the collision leads to the emergence  of an
oscillon at the collision center.
Notably, the regions of production of the solitons and the regions of the
parametric space where this process does not take place, are separated by a fractal-like boundary (Fig. \ref{fig:Fractal2}).
%
% \begin{figure}
%  \begin{center}
%    \includegraphics[height=0.75\columnwidth,angle=270]{snap3.eps}
%  \caption{\label{fig:1}\small Production of the kinks in the collision of two identical wave trains.
%  The initial and final field configurations are plotted at
%  $t=0$, $t=45$ and $t=100$ respectively.}
% \end{center}
% \end{figure}

For certain values of impact parameters an oscillon remaining in the collision center
decays into the second $K\bar K$ pair. Sometimes, two oscillons could also be ejected from the
collision center and after a while they could decay into two pairs of $K\bar K$. These observations confirm
the conclusion concerning the mechanism of the creation of the $K\bar K$ pair as a three-stage process.
In the first stage, the collision of the incoming excitations in the topologically trivial sector produces
an oscillon excitation. Next, the oscillon interacts with the incoming trains,  due to the
parametric resonance it may decay into the outgoing $K\bar K$ pair \cite{Romanczukiewicz:2010eg}.

These two creation mechanisms are in some way alike proving that the oscillon and the oscillating
mode can play similar roles in the dynamics.

\section{Kink-antikink scattering in the $\phi^6$ model.}
\label{Sec4}

Since the resonance scattering of the solitons was observed in several models with
different potentials \cite{Goodman:2005,CampbellPeyrard}, it was suggested
\cite{Goodman:2005} that the existence of an internal kink mode is a necessary
condition for the appearance of resonance windows. The parametrically modified sG model lent further
support to this view: depending on the value of a parameter,
kinks and anti-kinks do or do not possess an
internal mode; correspondingly, resonance windows do
or do not appear \cite{CampbellPeyrard}. Resonance windows have also
been observed in vector soliton collisions \cite{Goodman:2005a,Yang} and in
the scattering of kinks on impurities \cite{Kivshar:1991}. Again, the
mechanism always relies on the presence of a localized
internal mode, either of a single kink or of an impurity,
or both.

However, the mechanism of the reversible energy exchange may also work in the
absence of an internal oscillatory mode.
A new type of chaotic behavior, related with interplay between the states of perturbative spectrum and the kinks,
was observed recently in the $\phi^6$ model \cite{Dorey:2011yw}.
Although in this model the kink solution does not possess an internal vibrational mode, there still exist
multi-bounce resonance windows in the kink-antikink collisions. Thus, the mechanism of the energy transfer should
be different from the case of the $\phi^4$ model discussed above.

\begin{figure}
 \centering
\includegraphics[width=1\textwidth]{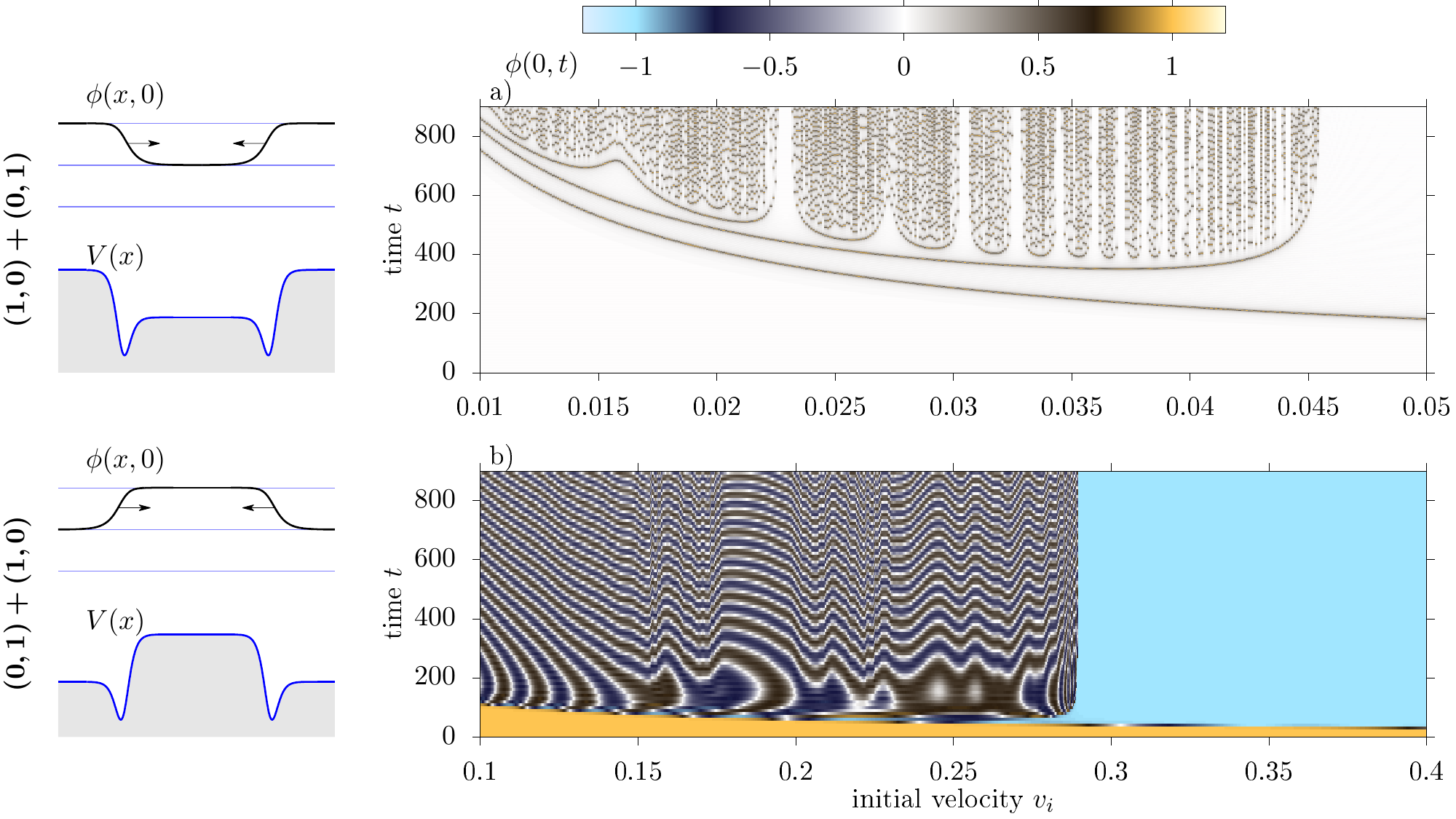}
   \caption{\small Two types of collisions of (identical) kinks in $\phi^6$ model (a) $(1,0)+(0,1)$ and (b) $(0,1)+(1,0)$.
   The left column shows the initial kink alignment and linearized potential, the right column the field at the center of collision $\phi(0,t)$ as
a function of initial velocity and time. Note the complicated fractal structure in the first case (a).}\label{fig:phi6frac}
\end{figure}

Recall that the $\phi^6$ model \re{Lag-phi6} has three vacua $\phi_v \in \{-1,0,1\}$. There are two different kink solutions
\re{kink-phi6}, so we have to consider two types of collision
of the solitons, the $K\bar K$ collisions in the vacuum sector $\phi_0=0$ and the collision in one of the symmetric
sectors  $\phi_0=\pm 1$.
In the former case the initial configuration of the colliding kinks, which we denote as
$(0,1)+(1,0)$ can be taken as a superposition
$\phi(x)=\phi_K(x+a) + \phi_{\bar K}(x-a)-1$ where $a$ is the separation parameter; in the latter case the initial
$K\bar K$ configuration $(1,0)+(0,1)$ is $\phi(x)=\phi_{K}(x-a) + \phi_{\bar K}(x+a)$.
Since both kinks do not posses an internal oscillating mode,  we can naively expect the collision will always be
quasi-elastic. However, there is a wide potential well in the case of widely separated $(1,0)+(0,1)$ $K\bar K$ pair
with two local minima associated with the positions of the solitons, see left column of Fig.~\ref{fig:phi6frac}.
In a contrast, in the case of $(0,1)+(1,0)$ configuration these two minima are separated by a barrier.
% \begin{figure}
%    \centering
%    \includegraphics[height=6cm,angle=0]{potentials.eps}
%    \caption{\label{fig:3}\small The potential $U(x)$ of the linear perturbations on the background of the composed
%    $(0,1)+(1,0)$ (up) and $(1,0)+(0,1)$ (bottom) $K\bar K$ configurations.}
% \end{figure}
%
% \begin{figure}
%    \centering
%    \includegraphics[width=1\textwidth]{phi6Potentials.pdf}
%    \caption{\label{fig:3}\small The potential $U(x)$ of the linear perturbations on the background of the composed
%    $(1,0)+(0,1)$ (left)  and $(0,1)+(1,0)$ (right) $K\bar K$ configurations.}
% \end{figure}
Thus, the energy of the colliding $(1,0)+(0,1)$ $K\bar K$ pair can be transferred into the excitation of the
trapped oscillating states of the composite configuration. With a suitable resonance condition, this
energy might be returned to the translational modes of the kinks allowing them to escape. On the contrary, there is no
oscillating modes in the collective potential of the $(0,1)+(1,0)$ $K\bar K$ pair, so the collision is expected to be elastic.

Indeed, numerical simulations confirm that the pattern of collision of the $(1,0)+(0,1)$ $K\bar K$ pair
is very similar to the picture observed in the $\phi^4$ model, see Fig.~\ref{fig:phi6frac}. As we can see on the plot (a)
an intricate pattern of the resonance escape windows exists up to a critical velocity $v_{cr}=0.0457$, after which the kinks always have enough
energy to separate.
Note that the impact velocities are very small, so the collision can be considered as an adiabatic process.

%
% \begin{figure}
%    \centering
%    \includegraphics[height=12cm,angle=0]{phi6.eps}
%    \caption{\label{fig:3}\small  $K\bar K$ collisions in the $(0,1)+(1,0)$ sector
% [plots (a), (b)] and in the $(1,0)+(0,1)$  sector [plots (c), (d), (e)]. Plots (a) and (c)
% show the fitted final velocity of the kink as a function of initial
% velocity, the plots (b) and (d) depict the field values at the collision center. The final plot, (e), shows the times
% to the first, second, and third kink-antikink collisions in the
% $(1,0)+(0,1)$ sector.}
% \end{figure}
By contrast,  the collision of the $(0,1)+(1,0)$ $K\bar K$ pair does not reveal any chaotic behavior with
a sequences of bouncing windows, for $v < v_{cr} \approx
0.289$ the pair annihilates into the vacuum $\phi_v=0$ with a small amount of radiation emitted, while for the $v > v_{cr}$
the collision yields a mirror pair of solitons escaping to infinity with no bouncing: $(0,1)+(1,0) \to (0,-1)+(-1,0)$.

By analogy with the $K\bar K$ scattering in the $\phi^4$ model, we can assign a ``bounce number" to each
resonance window in the $(1,0)+(0,1)$  sector. This is the number of collisions between the solitons  before
their final escape to infinity. The first two-bounce windows opens at $v_{in}\approx 0.0228$,
it is followed by a so called ``false" window at $v_{in}\approx 0.0273$, the kinks become well
separated after collision but they cannot escape and the  configuration finally collapses into the vacuum.
A third ``true" window opens at $v_{in}\approx 0.0303$, it is followed by a sequence of higher escape windows.
Furthermore, the regions near to the edges of the windows exhibit nested structures of higher bounce
windows revealing quasi-fractal structure of the interaction, which is similar to
the usual $\phi^4$ model. There is also the number of collective mode oscillations between two subsequent collisions,
for which numerical computations show that it corresponds to the excitation of
the lowest collective mode of the $K\bar K$ pair \cite{Dorey:2011yw}.

Thus, there exists a mechanism which allows resonances
to occur due to excitation of the collective oscillating states
in the potential well created in the space between the
constituents of a suitably ordered $K\bar K$  pair. This does not require the existence of an internal mode localized on a single kink.

Note that the initial velocity of the kinks is very small allowing
for adiabatic approximation in the analysis of fluctuations about the kink--anti-kink
configuration. Thus, the time between two subsequent collisions of the solitons is much larger than the corresponding period
in the $\phi^4$ model, and the number of internal oscillations of the kinks in the first window is quite large, $n=12$.
Further,  the effect of radiation pressure on the slow kinks is much stronger than in the $\phi^4$ model, it
affects the fine structure of higher order resonance windows \cite{Dorey:2011yw}.

\section{The role of quasinormal modes in topological defect collisions}
The internal oscillating modes, either attached to solitons or formed between them in the potential trap, play an
important role during the collisions.
They can live for a very long time thus, they  are the natural reservoirs storing the energy of perturbations
and giving it back to the
translational modes in a resonant way. The reversal energy exchange leads to the appearance of
fractal-like structures in the parameter space.
However, the internal oscillating modes are not the only modes where the energy can be stored \cite{Dorey:2011yw}.

Indeed, in the absence of the internal oscillating mode, usually the perturbations spread
rather quickly and only the modes with a slow group velocity near the mass threshold give
contribution to the long-time asymptotic relaxation with the usual power decay $\phi(0,t)\sim t^{-1/2}$.
Among these small perturbations there often are special modes which can stay exceptionally long near the potential center.
Such modes, which are referred to as the quasinormal modes (QNM) decay exponentially with time. They
play an important role in the relaxation processes in many physical
situations, such as, for example, collisions of black holes, radioactive decay or, in
or in acoustics, e.g. in the ringing of a bell \cite{MossBell}.

The QNM are often referred to as resonances, however to avoid possible confusion with the
resonant structure in the collisions of the solitons we do not use this term.
In more formal description these modes are often referred to as the poles of the Green function.
They appear as a result of modifications of the boundary conditions of
the corresponding linearized problem to purely outgoing wave. Such conditions break the Hermiticity  of the
linear operators, it results in appearance of the complex-valued
eigenfrequencies $\omega=\Omega+i\Gamma$.
The imaginary part $\Gamma>0$ is responsible for the exponential decay of the mode, $e^{i\omega t}=e^{-\Gamma t}e^{i\Omega t}$.
Note that neither $\phi^4$ nor sG models support QNMs due to  reflectionless nature of the
linearized potential, generated by the kinks. But this is an exception rather
than the rule.

Indeed, let us consider a small perturbation of the $\phi^4$ model with the deformed potential
\begin{equation}
 U(\phi,\epsilon) = W+\frac{m^2-4}{4}\frac{\epsilon W}{W+\epsilon}\, .
 \label{pot-def}
\end{equation}
Here $W=\frac{1}{2}(\phi^2-1)^2$ denotes the standard $\phi^4$ potential which is restored as the parameter $\epsilon$ vanishes.

\begin{figure}
 \begin{center}
 \includegraphics[width=0.75\linewidth]{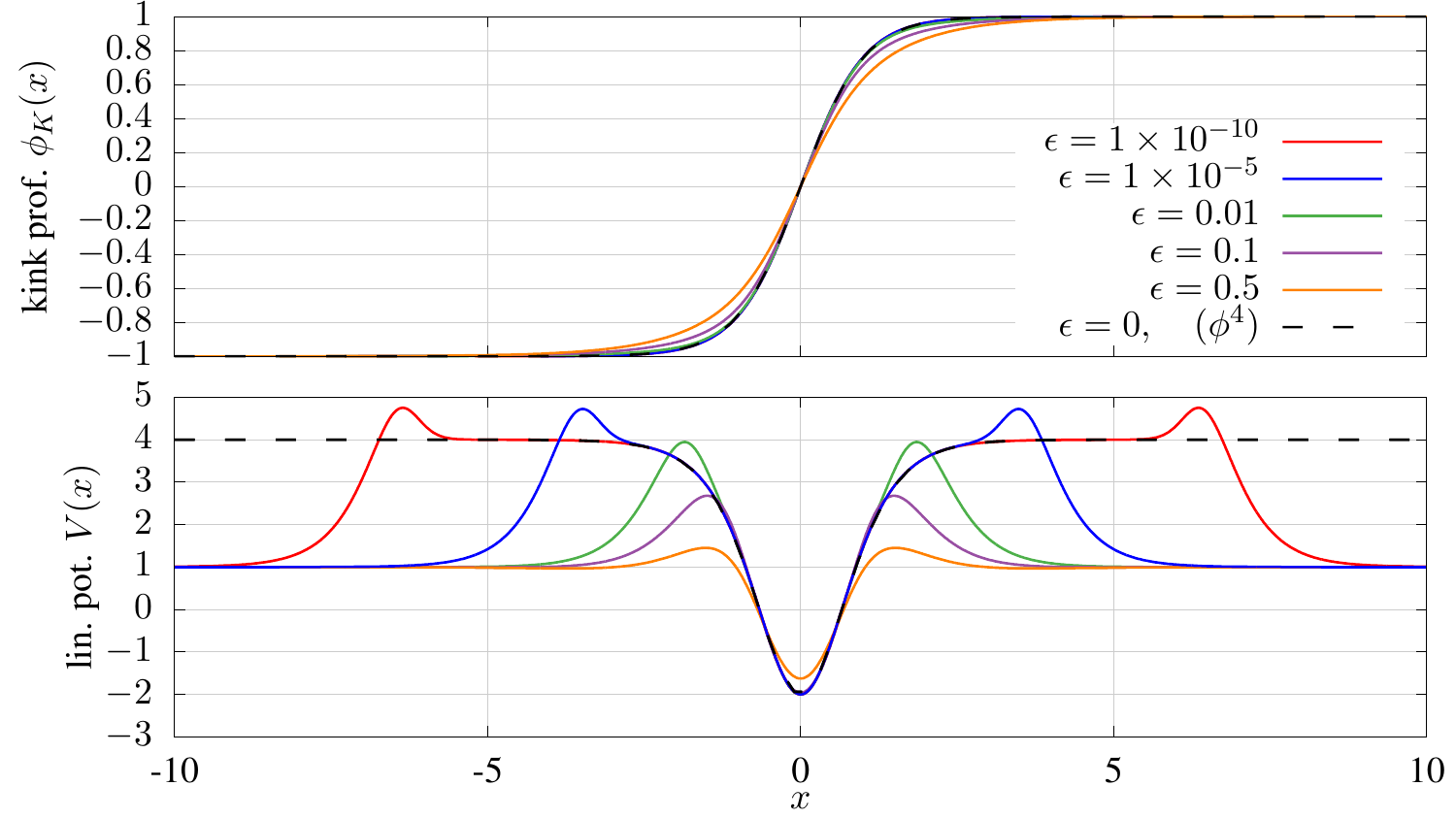}
 \caption{Kink profiles and the linearized potential $V(x) \re{pert-pot-def}$.
 Reprinted (without modification) \copyright2018 The Authors of \cite{Dorey:2017dsn}, Published by Elsevier
B.V.}\label{fig:kinksPotential}
\end{center}
\end{figure}

In the case of small positive values of $\epsilon$
the potential $ U(\phi,\epsilon)$ is almost the same as in the original
 $\phi^4$ theory, except the vicinity of the vacua. Near the vacua $W\approx 0$ and
$U(\phi,\epsilon)\approx \frac{m^2}{4}W$, which means that the mass of the small perturbations
is now equal to $m$ rather than $2$, as in the
standard $\phi^4$ model. The kink preserves its shape (Fig. \ref{fig:kinksPotential}), but it becomes
deformed as approaches the vacuum, $\phi-\phi_{vac}\sim e^{-mx}$.
Near the center the linearized potential
\begin{equation}
V(x)=\frac{\partial^2 U(\phi,\epsilon)}{\partial \phi^2} \,
\label{pert-pot-def}
\end{equation}
is similar to the  P\"oschl-Teller potential \re{perturb-phi4},
but asymptotically it tends to $V\to m^2$ instead of $2$, as in the original $\phi^4$ theory.

If $m<2$ the potential \re{pot-def} corresponds to a barrier of the width controlled by the parameter $\epsilon$ (See Fig.
\ref{fig:kinksPotential}).

\begin{figure}
 \begin{center}
 \includegraphics[width=0.75\linewidth]{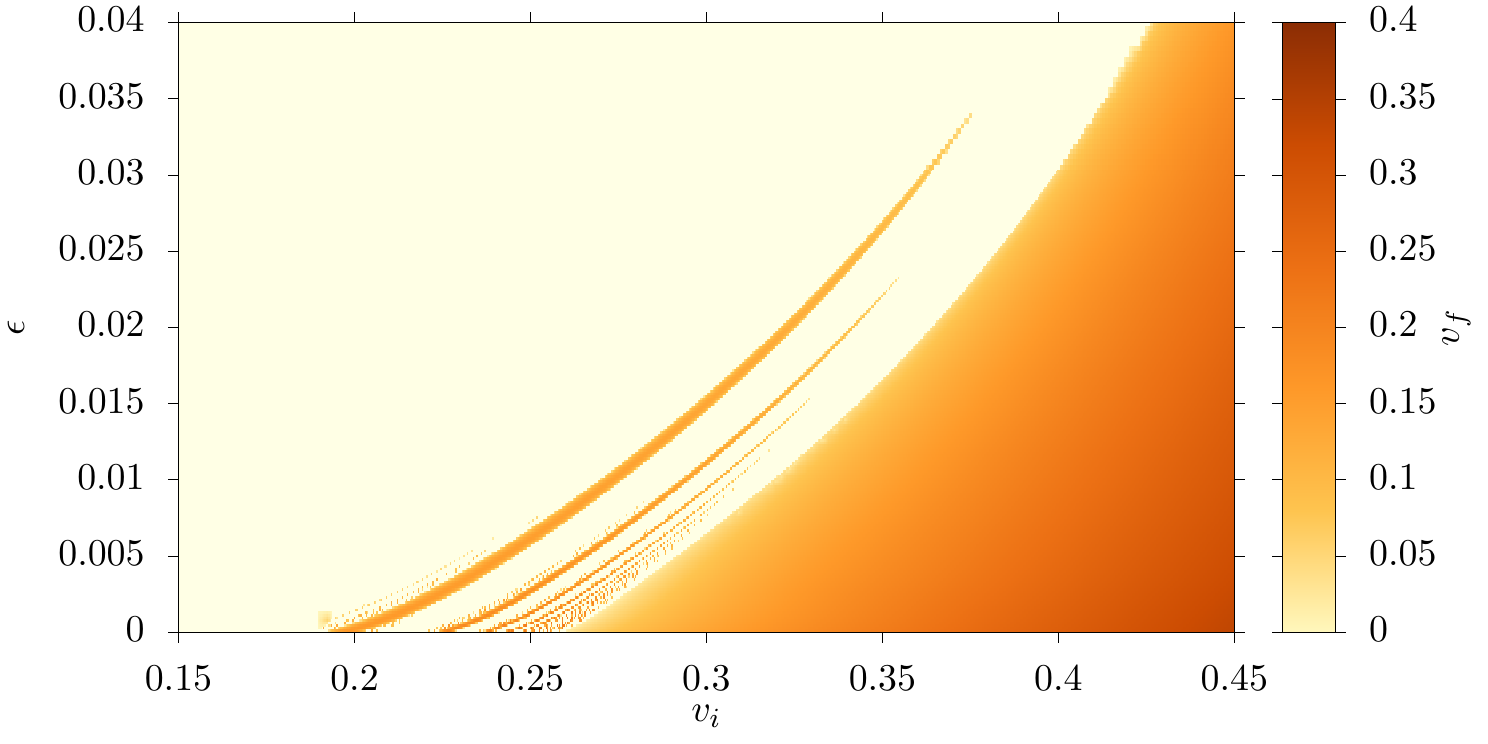}
 \caption{Final velocity $v_f$ of the kinks after the collision as a
 function of perturbation parameter $\epsilon$ and the initial
velocity $v_i$. Reprinted (without modification)  \copyright2018 The Authors of \cite{Dorey:2017dsn}, Published by Elsevier B.V.}\label{QNMCollisions}
\end{center}
\end{figure}

Let us consider the case  $m=1$. For small $\epsilon>0$ the linearized equation is almost
satisfied by the solution for the $\phi^4$ internal oscillating
mode localized by the potential \re{perturb-phi4}.
However since the potential $ U(\phi,\epsilon)$ asymptotically drops to $1$,
the frequency $\omega=\sqrt 3$ is moved to the continuum part of
the spectrum. The field is almost trapped near the soliton but now it can tunnel through the
potential barrier. Thus, the internal oscillating mode becomes the
quasi-normal mode. Note that its frequency for $\epsilon<0.04$ can be reasonably well (2\% accuracy) approximated  as
\begin{equation}
 \omega \approx 1.738 + 0.490\sqrt{\epsilon} -2.280\epsilon + (0.325\sqrt{\epsilon}+0.783\epsilon)i.
\end{equation}
Since almost all the properties of the model are continuously controlled by
the parameter $\epsilon$, we may also expect a continuous change of the properties of the modified model.
Although the QNM lose their energy due to the radiation more effectively than the standard internal
oscillating mode of the kink,
numerical simulations of the $K\bar
K$ collisions in the model with deformed potential \re{pot-def}
confirmed that the resonant structure remains visible for certain range of values of $\epsilon$ (see Fig.~\ref{QNMCollisions}).
The most narrow
windows closes for smaller values of $\epsilon$. The largest window close
at $\epsilon\approx0.034$. Another important result is that the critical
velocity for final escape grows with $\epsilon$. This is not a surprise since the lifetime of the QNM
decreases as $\epsilon$ increases. The energy
is stored for much shorter time. On the other hand, wide QNM (with large $\Gamma$) can be very effective
in transferring the excess of the energy away
from the solitons. This may be an important general mechanism responsible for binding solitonic configurations.

\section{Kink boundary scattering in the $\phi^4$ model.}
\label{Sec5}
Yet another example of give-and-take between the perturbation spectrum and solitons is provided by the
boundary $\phi^4$ model \re{Lag-phi} on semi-infinite line \cite{Dorey:2015sha}.
There are two vacua
$\phi_0 \in \{-1,+1\}$ on the left half-line $-\infty < x < 0$. The
bulk energy and Lagrangian densities are $\mathcal{E}=\mathcal{T}+\mathcal{V}$ and
$\mathcal{L}=\mathcal{T}-\mathcal{V}$ respectively, where
\begin{equation}
  \label{Lag}
\mathcal{T}=\frac12 \phi_t^2~~\mbox{and}~~
\mathcal{V}=\frac12 \phi_x^2+\frac12(\phi^2-1)^2\,.
\end{equation}
The boundary contribution to the energy is defined as $-H\phi_b$, where $\phi_b=\phi(0,t)$ and
$H$ can be interpreted as a boundary magnetic field. Thus,  the Neumann-type boundary
condition is $\phi_x(0,t)=H$ at $x=0$\,.

\begin{figure}
\centering
 \hspace*{-1pt}\includegraphics[width=0.8\textwidth,angle=0]{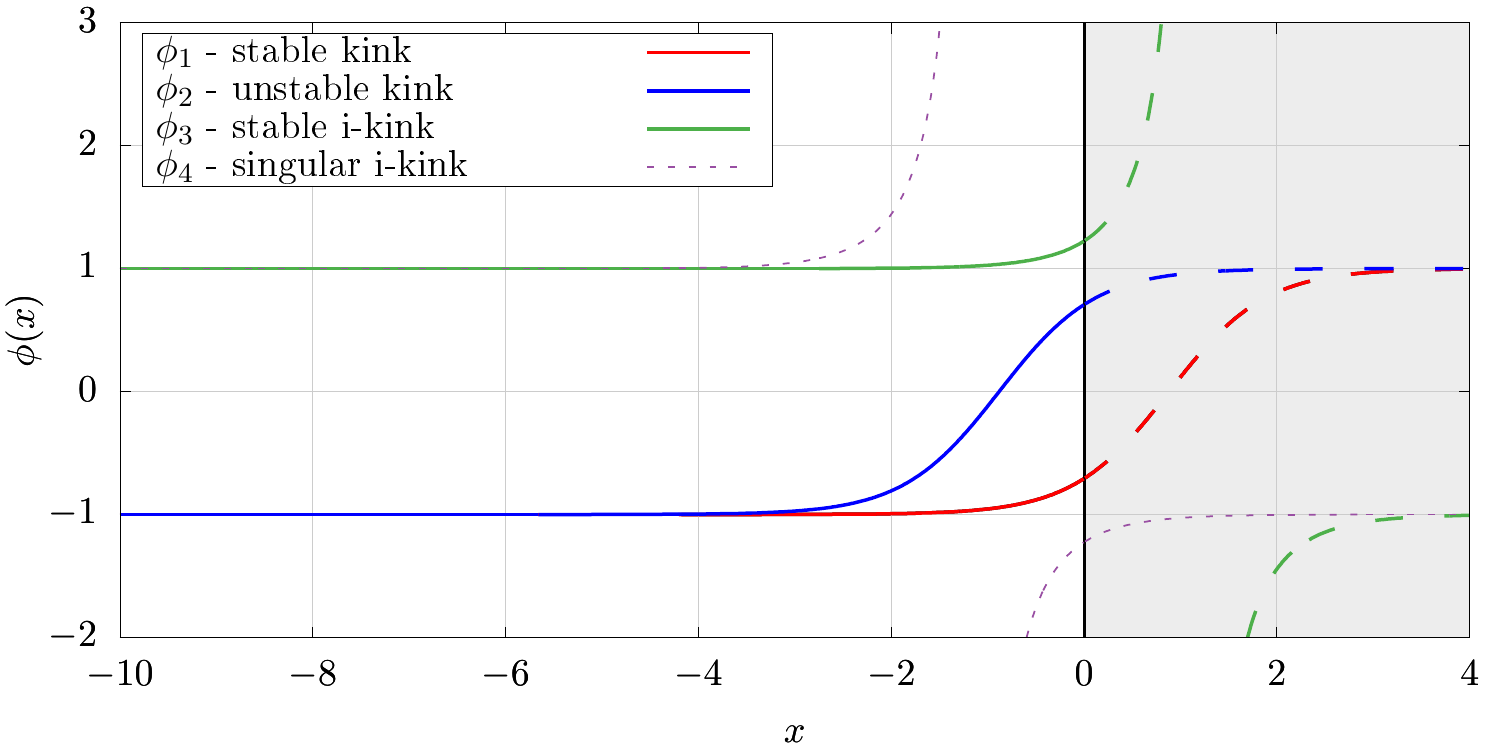}
 \vspace{-4pt}
 \caption{\label{fig:4}\small Static solutions for $H=1/2$.
Reprinted (without modification) from \cite{Dorey:2015sha}, $\copyright$
2015 The Authors of \cite{Dorey:2015sha}, under
the CC BY 4.0 license.}
 \vspace{-4pt}
\end{figure}

The states of the perturbative and non-perturbative spectrum of the boundary $\phi^4$ model are different
from those we discussed above in Section \ref{Sec1}. For $0 < H < 1$ there are four static solutions
of the field equations \cite{Dorey:2015sha}, see Fig.~\ref{fig:4}.
Two of them,
$\phi_1(x)=\tanh(x-X_0)$ and $\phi_2(x)=\tanh(x+X_0)$
with $X_0=\cosh^{-1}(1/\sqrt{|H|})$, are counterparts of the
kinks on the half-line, the other two solutions, $\phi_3(x)=-\coth(x-X_1)$ and
$\phi_4(x)=-\coth(x+X_1)$ with $X_1=\sinh^{-1}(1/\sqrt{|H|})$ are
irregular on the full line. On the
half line, $\phi_3$ is non-singular and corresponds to the
absolute minimum of the energy, while
$\phi_1$ is metastable, and $\phi_2$ is the unstable
saddle-point configuration between $\phi_3$ and $\phi_1$.

The energies of these configurations can be found by rewriting the functionals $E[\phi]
=\int_{-\infty}^0\mathcal{V}\,dx-H\phi_0$  as
\begin{equation}
E[\phi]=\frac{1}{2}\!\!\int\limits_{-\infty}^0 \!\!
\left(\phi_x\pm(\phi^2{-}1)\right)^2\! dx
\mp\left[\frac{1}{3}\phi^3{-}\phi\right]_{-\infty}^0
 - H\phi_0\,.
\label{BPS}
\end{equation}
Since $\phi_1$ and $\phi_2$ satisfy
$\phi_x=1-\phi^2$ we have
$\phi_1(0)=-\sqrt{1{-}H}$, $\phi_2(0)=\sqrt{1{-}H}$\,;
while
$(\phi_3)_x=\phi_3^2-1$ and so $\phi_3(0)=\sqrt{1{+}H}$.
This yields
\begin{align}
E[\phi_1]&= \frac{2}{3}-\frac{2}{3}(1{-}H)^{3/2}\,,~~~
E[\phi_2]= \frac{2}{3}+\frac{2}{3}(1{-}H)^{3/2}\,, \nonumber\\
E[\phi_3]&= \frac{2}{3}-\frac{2}{3}(1{+}H)^{3/2}\,.
\end{align}
The $\phi_3$ remains the only static solution for $H>1$. For negative values of the boundary field $H<0$,
physical solutions are $\tilde\phi_i(x)=-\phi_i(x)$, $i=1\dots 3$.

The states of the perturbative sector are also different from the corresponding excitations on the full line.
Considering the $\phi_1(x)$ as a static half-line solution with
$0<H<1$ in the $\phi(-\infty)=-1$ sector, we find the equation for linear perturbations $\xi(x)e^{i\omega t}$, with
boundary restriction $\partial_x \xi(x)=0$ at $x=0$,
\begin{equation}
\kappa^3-3\phi_0\kappa^2+(6\phi_0^2-4)\kappa-6\phi_0^3+6\phi_0=0
\label{eq:kappa}
\end{equation}
where $\phi_0=\phi_1(0)=-\sqrt{1-H}$, $k=i\kappa$ and now $\omega^2=4-\kappa^2$.
The solutions of (\ref{eq:kappa})  with $\kappa>0$ can give rise to the localized boundary
modes, the value of $\kappa$ must be less than mass threshold $2$ for the
corresponding $\omega$ to be real and the mode stable.
For $0<H<1$ there is just one positive solution
of Eq.~(\ref{eq:kappa}), which satisfies $\kappa<2$,
this is the single vibrational mode, localized near to the boundary.
The frequency of the mode changes from the mass threshold $m=2$ for $H=0$ to 0 for $H=1$.
For $H<0$ there is no such internal modes of the boundary, the spectrum of excitation is continuous.
Thus, the collision of the kinks with the boundary may lead to variety of resonance phenomena, which are similar
to the kink-anti-kink collisions in the $\phi^4$ theory on a full line.

However, there is another difference, related with the dynamics of the solitons near the boundary.
The static force between a single antikink and the boundary can be evaluated as \cite{Dorey:2015sha}:
$F = 32 \left( \frac{1}{4} H +e^{2x_0} \right)e^{2x_0} $ where $x_0<0$.
For $H<0$ the force is repulsive far from the boundary, and attractive in its vicinity.
Therefore, if the initial velocity $v_i$ is relatively small, the kink does not have
sufficient energy to overcome the initially-repulsive force,
it will be reflected without ever coming close to the boundary
and without exciting any other modes.

\begin{figure}
 \begin{center}
  \hspace*{-1pt} \includegraphics[width=0.75\textwidth,angle=0]{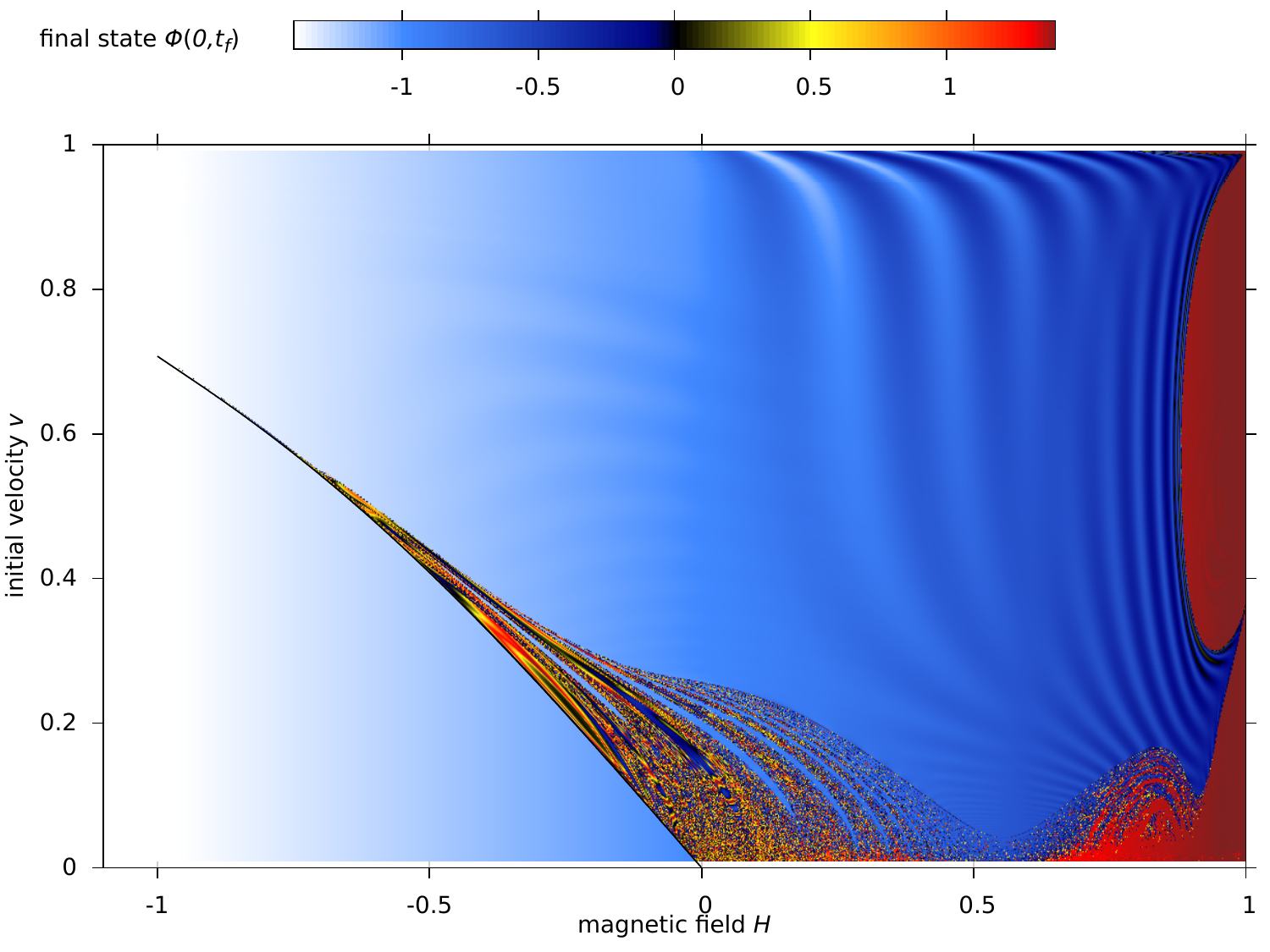}
 \vspace{-8pt}
 \caption{\label{fig:scan}\small A `phase diagram' of
antikink-boundary collisions. The plot
shows the value of the field at $x=0$,
as function of the initial velocity $v_{in}$ and the boundary magnetic field $H$.
Reprinted (without modification) from \cite{Dorey:2015sha}, \copyright
2015 The Authors of \cite{Dorey:2015sha}, under
the CC BY 4.0 license.}
\end{center}
 \vspace{-7pt}
\end{figure}

Indeed, numerical simulations revealed that for negative values of the boundary magnetic field $H$ and small
impact velocities $v_{in}$, the antikink
is reflected in a nearly elastic fashion, i.e., with very little
radiation, see Fig.\ref{fig:scan} (white and blue color indicates that the state of the boundary was not changed by the collision). As $v_{in}$ increases above some critical
value, the antikink becomes trapped by the boundary, leaving only radiation in the final state  (yellow and red).
Increasing $v_{in}$ further, scattering windows begin to open, until the initial velocity exceeds an upper
critical value and the antikink always escapes again. This pattern becomes more explicit for relatively small positive
values of $H$, see Fig.\ref{fig:scan}. Then the general structure of the resonant scattering windows remains similar
to the usual case of $K\bar K$ scattering in the model on a full line, it is exactly reproduced as $H=0$. However, there
is an important difference. For positive values of $H$ the initial translational energy
of the soliton colliding with the boundary can be stored not only in the internal
mode of the antikink, but also in the boundary mode.
Thus, the resonance condition for energy to be returned to the translational
mode of the antikink on a subsequent impact has to be modified. It leads to
the shifting  of the windows we observed in numerical simulations.

For larger positive values of the magnetic field $H$ new features appear. First,
the two-bounce window, which is ``missing'' in the full-line $K\bar K$ scattering \cite{Campbell:1983xu},
becomes resurrected. Secondly, as the magnitude
of the oscillation of the boundary mode becomes large enough, it can
be treated as an oscillon or kink-antikink bound state
trapped by the boundary. As we discussed in Section \ref{Sec3} above, this excitation can be considered
as an intermediate step in the process of production of the $K\bar K$ pair. However the general pattern of
non-perturbative dynamics of the solitons at $H\sim 1$ still remains obscure.

\section{Radiative decay of the internal oscillating mode}
The oscillational mode of the $\phi^4$ model in the linear approximation lasts forever and oscillates with constant amplitude and frequency.
However, the model is nonlinear and in the full model this statement is no longer true.
% Considering a perturbation series
% \begin{equation}
%  \phi(x,t)=\phi_K(x)+\sum_{n=1}^\infty A^n\xi^{(n)}(x,t)
% \end{equation}
% with $\xi^{(1)}=\eta_1\cos(\omega t)$ one can find that in the second order the equation takes the form
% \begin{equation}
%  \left(\partial_{tt}-\partial_{xx}+ V\right)\xi^{(2)}=-6\phi_K{\xi^{(1)}}^23\phi_K=-3\phi_K\eta_1^2\left(1+\cos(2\omega_1 t)\right)\,.
% \end{equation}
% Note that the source term on the rhs oscillates with the frequency $2\omega_1$. The solution of the above equation asymptotically has a form of
% outgoing wave
% \begin{equation}
%  A_{rad}=\frac{\pi q(q^2-1)}{32\sinh(\pi q/2)}\sqrt{\frac{q^2+4}{q^2+1}}\,,\qquad q=2\sqrt{\omega^2-1}\,,
% \end{equation}
% where $q$ corresponds to the wave vector for the outgoing wave frequency $2\omega_1$.
In the presence of quadratic terms in equations of motion, a  mode oscillating with the frequency $\omega_1$ becomes a source of
radiation with frequency $2\omega_1$ since $\cos^2(\omega_1t)=\frac{1}{2}(\cos(2\omega_1)+1)$.
The radiation  carries away the energy from the mode causing its decay.
The rate in which the energy is radiated out can be found from the energy flux of the outgoing wave  $dE/dt\sim\phi_x\phi_t\sim -A^4$.
Since the energy stored in the internal
oscillating mode is proportional to $A^2$ this leads to the relation (a Manton-Merabet power law
\cite{Manton:1996ex})
\begin{equation}
 \frac{dE}{dt}\sim -A\frac{dA}{dt}\sim -A^4\Rightarrow \frac{dA}{dt}\sim-A^3\Rightarrow A\sim t^{-1/2}\,.
\end{equation}
An additional  consequence of the nonlinearities is that the apparent frequency of the solution changes and, in general, is lowered
$\omega(A)=\omega_1-A^2\omega^{(2)}$. A very similar effect to the slowing down of the pendulum for large amplitudes.
When the mode is highly excited, instead of the slow decay, a pair of a kink-antikink structures
can be ejected transforming the original kink into an antikink.

\begin{figure}
 \begin{center}
  \hspace*{-1pt} \includegraphics[width=1\textwidth,angle=0]{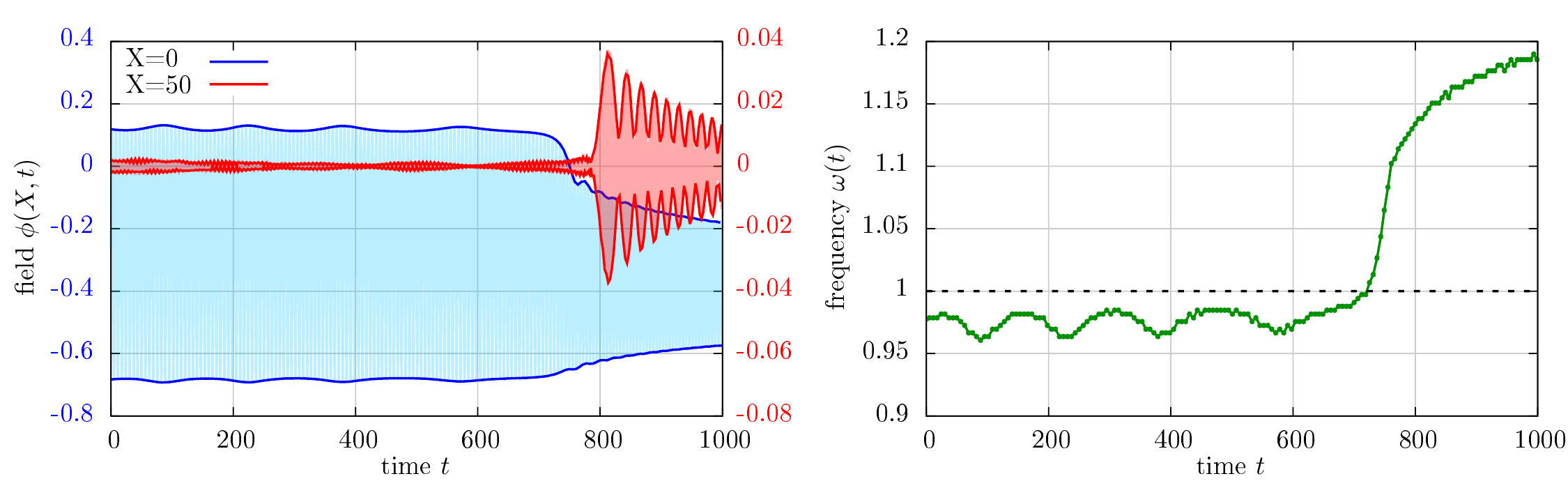}
 \vspace{-8pt}
 \caption{\label{fig:slowfast}\small Transition from slower to faster decay for $H=0.8393$ and $A_0=0.3$. On the left plot the blue line shows the
field at center and red the far field. Right panel shows the frequency measured from the two subsequent maxima. Clearly the transition occurs when
the frequency crosses the value $m/2=1$ at $t\approx 720$. }
\end{center}
 \vspace{-7pt}
\end{figure}

The above result, $A\sim t^{-1/2}$, is true when the internal oscillating mode decays through the radiation of the second harmonic.
In the $\phi^4$ model the condition is fulfilled. However, in general this does not need to be true.
The boundary described in the previous section has an oscillating mode, frequency of which can be controlled by the value of $H$ in the range from
0 to 2. If the oscillating mode has the frequency below $m/2=1$, the second harmonics is unable to propagate.
Higher orders of the perturbation series would have to be considered.
If the frequency of the oscillating mode would be between $m/3$ and $m/2$ the first propagating harmonic is the third.
For lower frequencies even that is not enough and higher and higher harmonics have to be considered.
In general, if the first harmonic, which can propagate, is enumerated $n$, the Manton-Merabet changes to
\begin{equation}
 A\sim t^{-1/(2n-2)}
\end{equation}
Because the amplitude affects the frequency of the mode, it is possible that for certain amplitude the frequency $\omega(A)=\omega_1-A^2\omega^{(2)}$
would be below $m/2$ even for $\omega_1>m/2$.
However, as the mode decays slowly $A\sim t^{-1/4}$, its amplitude decreases and frequency grows crossing at some point $m/2$.
This event releases the second harmonics and the decay rate increases to $A\sim t^{-1/2}$.
Far away from the mode this transition can be seen as a sudden burst of radiation. An example of such an evolution is presented in the Figure
\ref{fig:slowfast}.
For $H=0.8393$ the eigenfrequency of the internal oscillating mode is equal to $\omega_1=1.2467$.
A highly excited mode, with initial amplitude  $A_0=0.3$ is slowed down by the nonlinearities and the observed frequency $\omega(A)\approx0.98$ is
below half of the threshold $m/2=1$. The mode decays through the third harmonic exhibiting rather slow decrease of the amplitude and small radiation
in the far field zone. Around $t=720$ the amplitude decreased to the value for which the frequency becomes $m/2=1$. The second harmonic is freed and
the decay rate accelerates. At certain distance this event can be measured as a burst of radiation.

\section{Conclusions}
In this brief review we presented some recent results in investigation of interactions between the kinks,
related with fine interplay between the solitons and excitations of the perturbative spectrum.
We demonstrate that the spectral structure of small perturbations around the kinks plays an essential role in the dynamics of the solitons.
Apart from the well known mechanism of resonant collisions in $\phi^4$ explained by the existence of the internal
oscillating mode of the kink, we
have shown that other resonant mechanism can exist as well. The modes trapped between asymmetric kinks
in $\phi^6$, or the quasinormal modes, may play the same role as the internal mode of a kink.

We have also discussed a reverse process, when the kinks are created from the radiation. In this resonance
process the oscillons and the internal modes of the kinks may play a role of
an intermediate state, increasing the excess of energy density which is released by the
emission of kink-antikink pairs.

Another interesting problem, which we briefly sketched, is the similarity between the internal oscillating modes
and the oscillons. In some ways they can be
transformed from one into another via adiabatic change of the model parameters.

We show that the interaction between the solitons and incoming radiation may produce some unexpected results.
Indeed, in the integrable sine-Gordon model, the waves pass through the kink without
any exchange of energy. Similar effect may be observed in the $\phi^4$ model only if we restrict ourselves to the linear
approximation. The contribution of higher nonlinear terms yields both the
the energy and momentum transfer between the kink and the second harmonic. Furthermore, the contribution of the
second harmonic produces a surplus of the momentum
behind the kink pushing it towards the source of the radiation. This is an example of the negative radiation pressure effect.

In the case of the asymmetric kinks in the $\phi^6$  model the force exerted by
the waves is also asymmetric. On one side of the soliton the wave causes positive
radiation pressure, however, on the other side of the kink, the radiation pressure is negative. As a result,
any small perturbation pushes the kink in such a way, that the vacuum with smaller mass
parameter always expands.
%This effect leads to a chain reaction of kink-antykink collisions.

\section*{Acknowledgments}
We are grateful to Patrick Dorey for valuable collaboration, many results of our joint work are reviewed in this brief survey.
We thank Piotr Bizo\'n and for inspiring and valuable discussions.
Y.S. gratefully acknowledges support from the  JINR Bogoljubov-Infeld Program of collaboration Krak\'ow-Dubna. He
would like to thank the Institute of Physics, Jagiellonian University Krak\'ow, for its kind hospitality.

%%%%%%%%%%%%%%%%%%%%%%%%%%%%%%%%%%%%%%%%%%%%%%%%%%%%%%%%%%%%%%%%%%%%%%
\bibliographystyle{ieeetr}
\end{document}